\documentclass[iop]{emulateapj}
 
\usepackage{wasysym}
\usepackage{natbib}
\usepackage{longtable}
\usepackage{graphicx}
\usepackage{fancyref}
\usepackage{gensymb}
\usepackage{adjustbox}
\usepackage{booktabs}
\usepackage{scrextend}

\newcommand{\ra}[1]{\renewcommand{\arraystretch}{#1}}

\shortauthors{Fischer et al.}

\begin{document}

\title{Spatially Resolved Outflows In a Seyfert Galaxy at z = 2.39}

\author{Travis C. Fischer\altaffilmark{1,2}\altaffilmark{\textdagger},
J. R. Rigby\altaffilmark{1}, 
G. Mahler\altaffilmark{3}, 
M. Gladders\altaffilmark{4,5}, 
K. Sharon\altaffilmark{3}, 
M. Florian\altaffilmark{1}, 
S. Kraemer\altaffilmark{2},
M. Bayliss\altaffilmark{6}, 
H. Dahle\altaffilmark{7}, 
L. Felipe Barrientos\altaffilmark{8}, 
S. Lopez\altaffilmark{9},
N. Tejos\altaffilmark{10},
T. Johnson\altaffilmark{3}, 
E. Wuyts\altaffilmark{11}}

\altaffiltext{1}{Observational Cosmology Lab, Goddard Space Flight Center, Code 665, Greenbelt, MD 20771, USA}

\altaffiltext{2}{Institute for Astrophysics and Computational Sciences, Department of Physics, The Catholic University of America, Washington, DC 20064, USA}

\altaffiltext{3}{Department of Astronomy, University of Michigan, 500 Church Street, Ann Arbor, MI 48109, USA}

\altaffiltext{4}{Department of Astronomy \& Astrophysics, University of Chicago, 5640 South Ellis Avenue, Chicago, IL 60637, USA}

\altaffiltext{5}{Kavli Institute for Cosmological Physics, University of Chicago, 5640 South Ellis Avenue, Chicago, IL 60637, USA}

\altaffiltext{6}{MIT Kavli Institute for Astrophysics and Space Research, 77 Massachusetts Avenue, Cambridge, MA 02139, USA}

\altaffiltext{7}{Institute of Theoretical Astrophysics, University of Oslo, P.O. Box 1029, Blindern, NO-0315 Oslo, Norway}

\altaffiltext{8}{Instituto de Astrofisica, Pontifica Universidad Catolica de Chile, Vicuna Mackenna 4890, Santiago, Chile} 

\altaffiltext{9}{Departamento de Astronom\'ia, Universidad de Chile, Casilla 36-D, Santiago, Chile}

\altaffiltext{10}{Instituto de F\'isica, Pontificia Universidad Cat\'olica de Valpara\'iso, Casilla 4059, Valpara\'iso, Chile}

\altaffiltext{11}{ArmenTeKort, Antwerp, Belgium}

\altaffiltext{\textdagger}{James Webb Space Telescope NASA Postdoctoral Program Fellow; travis.c.fischer@nasa.gov}

\begin{abstract}

We present the first spatially resolved analysis of rest-frame optical and UV imaging and spectroscopy for a lensed 
galaxy at z = 2.39 hosting a Seyfert active galactic nucleus (AGN). Proximity to a natural guide star has enabled high 
signal-to-noise VLT SINFONI + adaptive optics observations of rest-frame optical diagnostic emission lines, which exhibit an 
underlying broad component with FWHM $\sim$ 700 km/s in both the Balmer and forbidden lines. Measured line ratios place 
the outflow robustly in the region of the ionization diagnostic diagrams associated with AGN. This unique opportunity 
--- combining gravitational lensing, AO guiding, redshift, and AGN activity --- allows for a magnified view of two main 
tracers of the physical conditions and structure of the interstellar medium in a star-forming galaxy hosting a weak AGN 
at cosmic noon. By analyzing the spatial extent and morphology of the Ly$\alpha$ and dust-corrected H$\alpha$ emission, 
disentangling the effects of star formation and AGN ionization on each tracer, and comparing the AGN induced mass outflow 
rate to the host star formation rate, we find that the AGN does not significantly impact the star formation within its 
host galaxy.

\end{abstract}


~~~~~

\section{Introduction}
\label{sec1}

Galaxies at the peak of cosmic star formation live in a state of punctuated equilibrium, where continuous accretion 
of gas from the cosmic web feeds large molecular gas reservoirs, and is balanced by star formation and 
outflows. Galactic wind feedback is widely acknowledged to play a critical role in the evolution of galaxies 
by expelling gas from the central regions of galaxies, shutting down their global star formation and 
regulating their stellar mass and size growth \citep{Dav12,Vog13}. However, 
the physical mechanisms involved and the relative importance of active galactic nuclei (AGN) and star 
formation as the main feedback drivers remain poorly understood. AGN-driven feedback is evident in 
luminous but rare QSOs and radio galaxies, but observational evidence is lacking for AGN feedback in 
less extreme, $“$normal$”$ star-forming galaxies \citep{Fab12}. In optical and infrared spectroscopy, evidence 
of AGN outflows, which can produce feedback, are observed as relatively broad emission-lines, with full 
width at half maximum (FWHM) $>$ 250\,km s$^{-1}$, inside the Narrow Line Region (NLR), a region of relatively 
low density ionized gas extending from the nuclear torus to distances of hundreds to thousands of parsecs from 
the nucleus.

\begin{figure*}[t]
\centering

\includegraphics[width=0.95\textwidth]{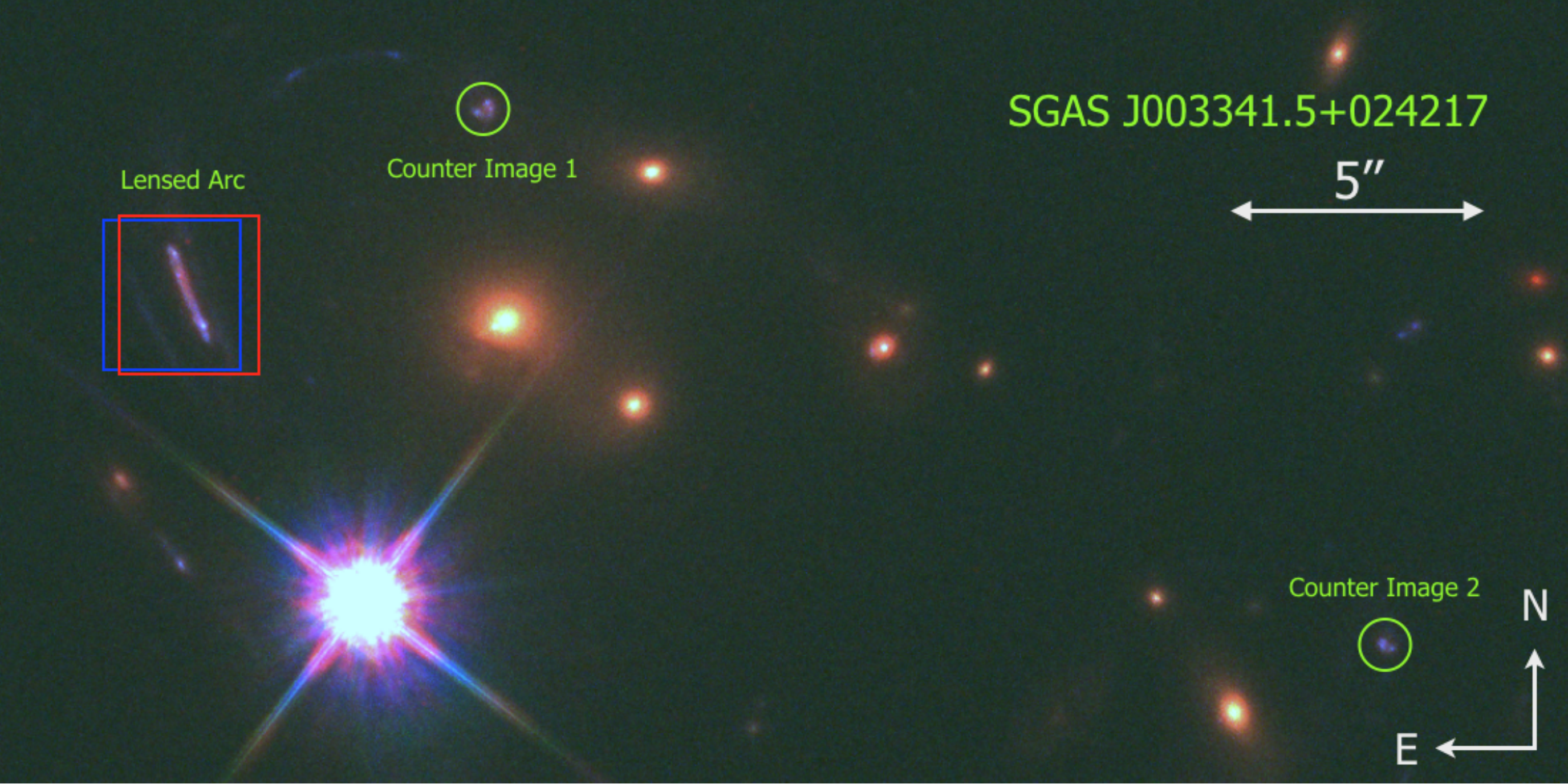}

\caption{{\it HST}/WFC3 F555W/F814W/F105W composite image of the lensed galaxy SGAS J003341.5+024217. The galaxy is 
multiply imaged as an elongated arc and two counter images. Fields of view for SINFONI H- and K-band observations of 
the arc are represented by blue and red boxes, respectively. The bright source to the southeast is a star, which was 
used as the natural guide star for SINFONI adaptive optics.}
\label{fig:field}

\end{figure*}

Recent studies by \citet{Fis17,Fis18} find outflows may not be powerful 
enough in nearby AGN to drive gas out to bulge-radius distances of 2 -- 3 kpc. Kinematics within the 
NLR are largely due to rotation and in situ acceleration of material originating in the host 
disk. Spatially-resolved outflowing gas in Type 2 Seyferts and nearby (z $<$ 0.12) QSO2s 
extends to a fraction of radii typical of host galaxy stellar bulges (r $\sim$ 2 -- 3 kpc). 
These findings suggest that outflows at z $\sim$ 0 may not be powerful enough to evacuate gas from 
their entire bulges. Several other studies have reached similar conclusions \citep{Kar16,Vil16,Kee17,Ram17}. 

\citet{For14} and \citet{Gen14} have reported evidence for likely AGN-driven outflows 
in the central regions of massive (log(M$_*$/M$_{\astrosun}$)) $\geq$ 10.9) main-sequence star-forming galaxies 
(SFGs) at high redshifts (z $\sim$ 2) with FWHM $\sim$ 1000 -- 1500\,km s$^{-1}$ and elevated [N II]/H$\alpha$ 
ratios $\geq$ 0.5. The outflows are resolved over the inner 2 -- 3 kpc of the galaxies and detected in the 
forbidden [N~II] and [S~II] lines as well as in H$\alpha$. Therefore, these broad emission lines can not be 
due only to a virialized, parsec-scale AGN broad-line region. The mass outflow rates are estimated to be comparable 
to or exceed the star-formation rate (SFR) of the galaxy, thus creating an important avenue for the quenching of 
star formation. The next step is to measure the size, geometry, velocity profile and mass loading through high-resolution 
mapping of an outflow region. However, sensitivity and spatial resolution restrictions currently limit us to 
barely resolving ionized-gas structures in only a few of the largest and most massive SFGs at z $\sim$ 2.

Our team has recently discovered a bright, lensed galaxy, SGAS J003341.5+024217, henceforth SGAS 0033+02 
(Figure \ref{fig:field}), as described in the Magellan Evolution of Galaxies Spectroscopic and Ultraviolet 
Reference Atlas (MegaSaura; \citealt{Rig18}), that offers a unique opportunity to spatially resolve the 
influence of AGN feedback in a galaxy residing near Cosmic Noon at z $\sim$ 2.4.

SGAS 0033+02 was identified as a candidate lensed system through the Sloan 
Giant Arcs Survey (Gladders et al. in prep) in which objects with arc-like 
morphology are identified along lines of sight with photometric evidence 
for cluster- or group-scale masses, via a direct visual examination of 
Sloan Digital Sky Survey imaging data. Follow-up $gri$ imaging acquired 
with the MOSCA imager on the 2.5m Nordic Optical Telescope on UT 15 
September 2012 confirmed the arc-like morphology of this system, and a 
spectroscopic redshift of z=2.378 was obtained with the same telescope 
using the ALFOSC spectrograph on the Nordic Optical Telescope on UT 01 
September 2013.


Fortuitously, a bright (g$\sim$15.4) star appears in 
projection only 7\arcsec\ from the main image of the lensed arc 
SGAS 0033+02.  Recognizing this, we obtained laser guide star adaptive 
optics observations with SINFONI instrument on the VLT 

VLT/SINFONI IFU observations of outflows in luminous 1.5<z<3 AGN have been resolved in detail in previous 
studies \citep{Nes08,Nes11,Per15,Cres15,Car15,Bru16,Nes17,For18}. However, through the 
combination of observations across several observatories, we are able to spatially resolve the size, geometry, 
and mass loading of AGN outflows on scales of 10s of parsecs for the first time at high redshift.

\begin{figure*}[t]
\centering
\includegraphics[width=0.52\textwidth]{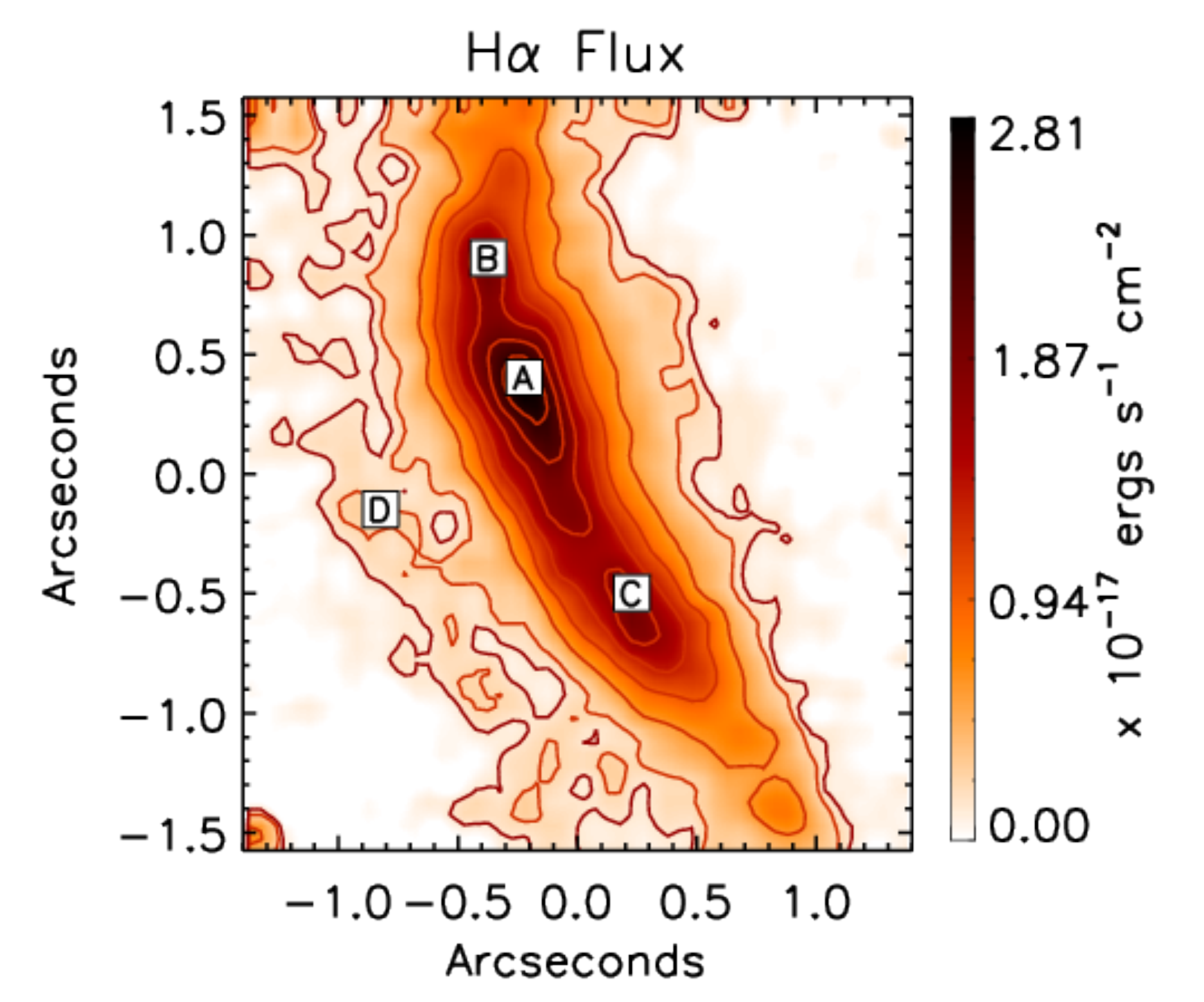}
\includegraphics[width=0.46\textwidth]{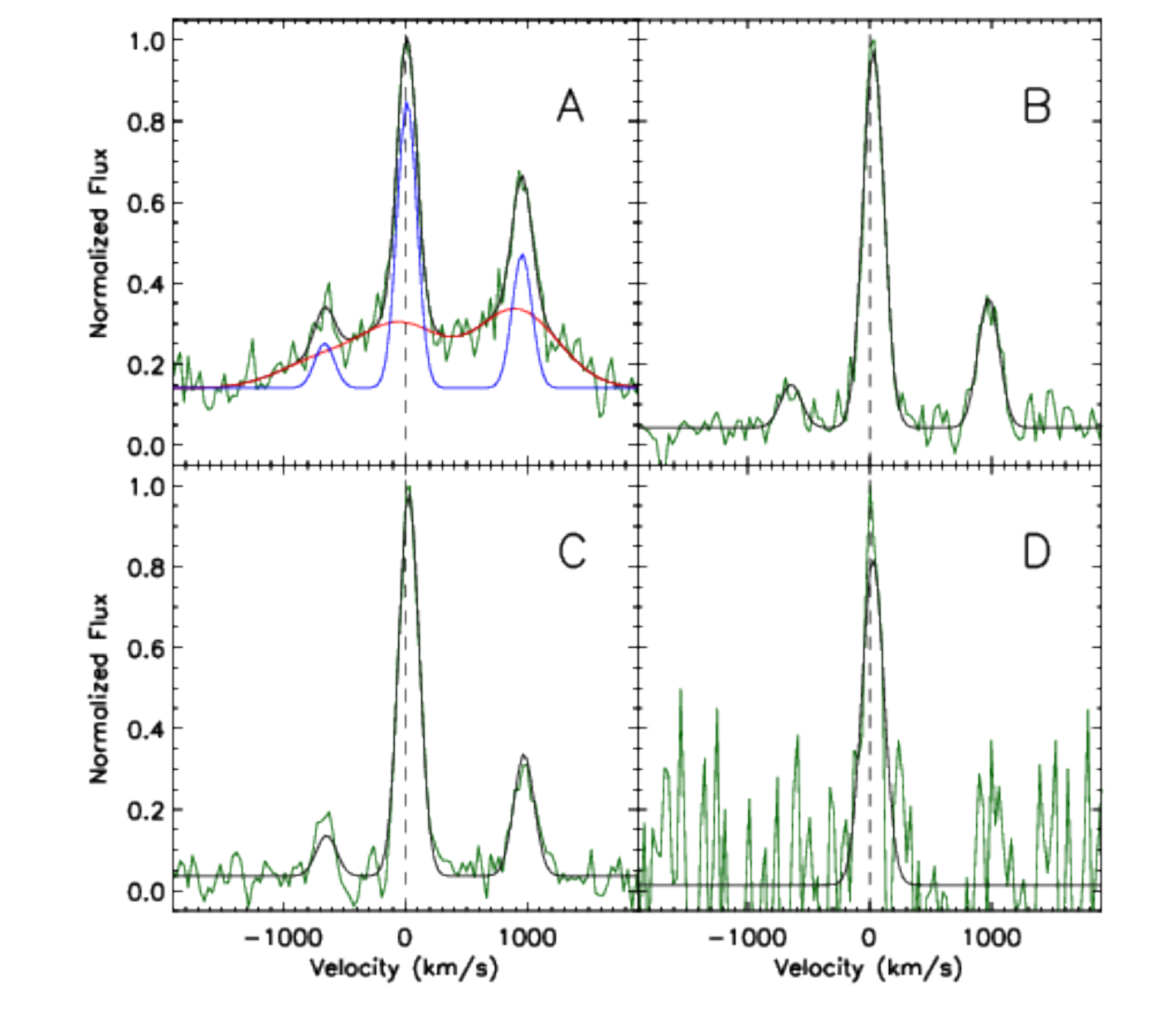}

\caption{Left: Continuum-subtracted H$\alpha$ flux distribution in SGAS 0033$+$2 obtained from the combined VLT/SINFONI $\sim3'' \times 3''$ K-band IFU datacube, smoothed with a Gaussian of FWHM = 2.0 x 2.0 pixels. Outer, dark red contours represent a 3$\sigma$ S/N lower flux limit. Right: Spectra of individual spaxels sampling four H$\alpha$ regions, each over plotted with their best fitting model, with approximate spaxel positions labeled in the H$\alpha$ image. Solid black line represents the total model. Blue and red lines represent H$\alpha~ +$ [N II] $\lambda\lambda$ 6548,6583 narrow and wide Gaussian components, respectively. Vertical
dashed black line represents the H$\alpha ~ \lambda$6563 wavelength at systemic velocity.}
\label{fig:spectra}

\end{figure*}

\section{Observations and Data Reduction}
\label{obs_sec}

\subsection{HST WFC3 Imaging Observations}

Imaging of SGAS 0033+02 was acquired using the HST Wide Field Camera 3 during two visits on 2016 October 30 and 2016 November 8. 
In the IR channel, images were taken in the F140W and F105W filters with cumulative exposure times of 459s and 1026s respectively.  
In the UVIS channel, exposures were taken in the F410M, F814W, and F555W filters with cumulative exposure times of 7256s, 1900s, 
and 1748s respectively.  At the redshift of the source, z = 2.39, these filters provide a wide wavelength coverage, but isolate 
Ly$\alpha$ emission entirely within the F410M filter.

The HST imaging data were reduced using the software package, DrizzlePac\footnote[1]{drizzlepac.stsci.edu}. Images were aligned 
using tweakreg, then drizzled, using astrodrizzle, to a common reference grid with a scale of 0.03 arcseconds/pixel, 
with a Gaussian kernel and a drop size of 0.8. Three hot pixels in the IR channel near or within the main arc consistently 
failed to flag in astrodrizzle, resulting in artifacts in the final data products that could easily be mistaken for real 
substructure within the arc. These hot pixels were flagged manually in the data quality extension of the flat-field calibrated 
files before creating the final drizzled images, creating final data products free from these artifacts.

Continuum-subtracted Ly$\alpha$ imaging was produced using the F410M medium band filter with the F555W filter providing 
the continuum flux. Given the high equivalent width of Ly$\alpha$ in the MagE spectrum described below, EW$_{obs}=$203\,\AA, and 
the F410M bandpass of 70\,\AA, we calculate that Ly$\alpha$ contributes 74$\%$ of the flux in F410M, with the remainder 
coming from continuum. We then scale the F555W image to match that continuum level, using annular aperture photometry 
of SGAS0033 in the F410M and F555W {\it HST} images, covering the same region as the MagE aperture.

\subsection{MagE Magellan Observations}
Observations of SGAS 0033+02 were obtained with the MagE instrument on the Magellan Baade telescope in UT 2015 Nov 07 and 10, 
for a total of 7 hrs of integration.  The spectra cover observed wavelengths of 3200 - 8280\AA, including Lyman alpha.  Description 
of the observations and data reduction, and the MagE spectra themselves, were published by \cite{Rig18}.  Their Figure 1 
shows that over the course of the observations, the 2”x10” MagE slit covered the full extent of the SGAS 0033+02 arc.

\subsection{VLT SINFONI and MUSE IFU Observations}

Observations of SGAS 0033+02 using VLT/SINFONI+AO were taken across several nights (2015 September 8, October 10, and 
December 4, 6, 9, and 12) in the H-, and K-bands, with resolving powers of R = $\lambda$/$\delta\lambda$ = 3000 
and 4000 and covering spectral regions between 1.45 - 1.85 and 1.95 - 2.45 $\mu$m respectively, with a pixel scale 
0.05$" \times$ 0.1$"$ and sampling a field of view of 3.2$" \times$ 3.2$"$. Observations were carried out in observing 
blocks (OBs) of an OSOOSO pattern, alternating object (O) and sky (S) positions. Each OB was dithered by 0.15$''$ around 
the central position to mitigate bad pixels and cosmic rays. 8 individual exposures of 600s were obtained in the 
H-band and 28 individual exposures of 600s were obtained in the K-band, for totals of 1 hr 20 min and 4 hr 40 min 
of on-source integration, respectively. VLT/SINFONI data were reduced using the software package SPRED developed 
specifically for SPIFFI \citep{Sch04,Abu06} following the procedures described in \citet{For09}. The offsets 
between individual cubes were determined from the known dither pattern within each OB, and the location of the 
acquisition star observed before each OB. The final PSF is created by fitting a circularly symmetric 2D Gaussian profile to 
acquisition star exposures taken prior to each OB of the science target, and results in a FWHM of 0.19$”$ in K and 
0.18$”$ in H-band. The PSF FWHMs correspond to the effective resolution of all observations for our target. 
Early B-type standard stars were observed each night to provide flux calibration and telluric correction.

Observations of SGAS 0033+02 using VLT/MUSE were obtained under the program 098.A-0459(A). The 1 arcmin field-of-view is 
sampled with 349 $\times$ 352 0.2$''$ wide spaxels. Our setup provided a wavelength range from 4650 to 9300 \AA ~at a resolving 
power R ranging from 2000 to 4000. Each spectral bin is 1.25 \AA~ wide. The observations were carried out in ’service mode’ 
during dark time, with clear-sky conditions, airmass below 1.8, and seeing better than 0.7$''$ on the nights of 2017 September 19, 20. We obtained a total of 12 exposures of 700s on-target each. The exposures were taken 
within “Observing Blocks” of 4 exposures each. We applied a small dithering and 90 deg rotations between exposures to 
reject cosmic rays and minimize patterns of the slicers on the processed combined cube. We reduced all the observations 
using the MUSE pipeline recipe v1.6.4 and ESO reflex v2.8.5. The individual exposures were combined into one final 
science datacube. The total on-target time was therefore 2.3 hours. The sky subtraction was improved on this cube 
using the Zurich Atmospheric Purge (ZAP) algorithm v1.0.

\begin{figure*}[h]
\centering

\includegraphics[width=0.31\textwidth]{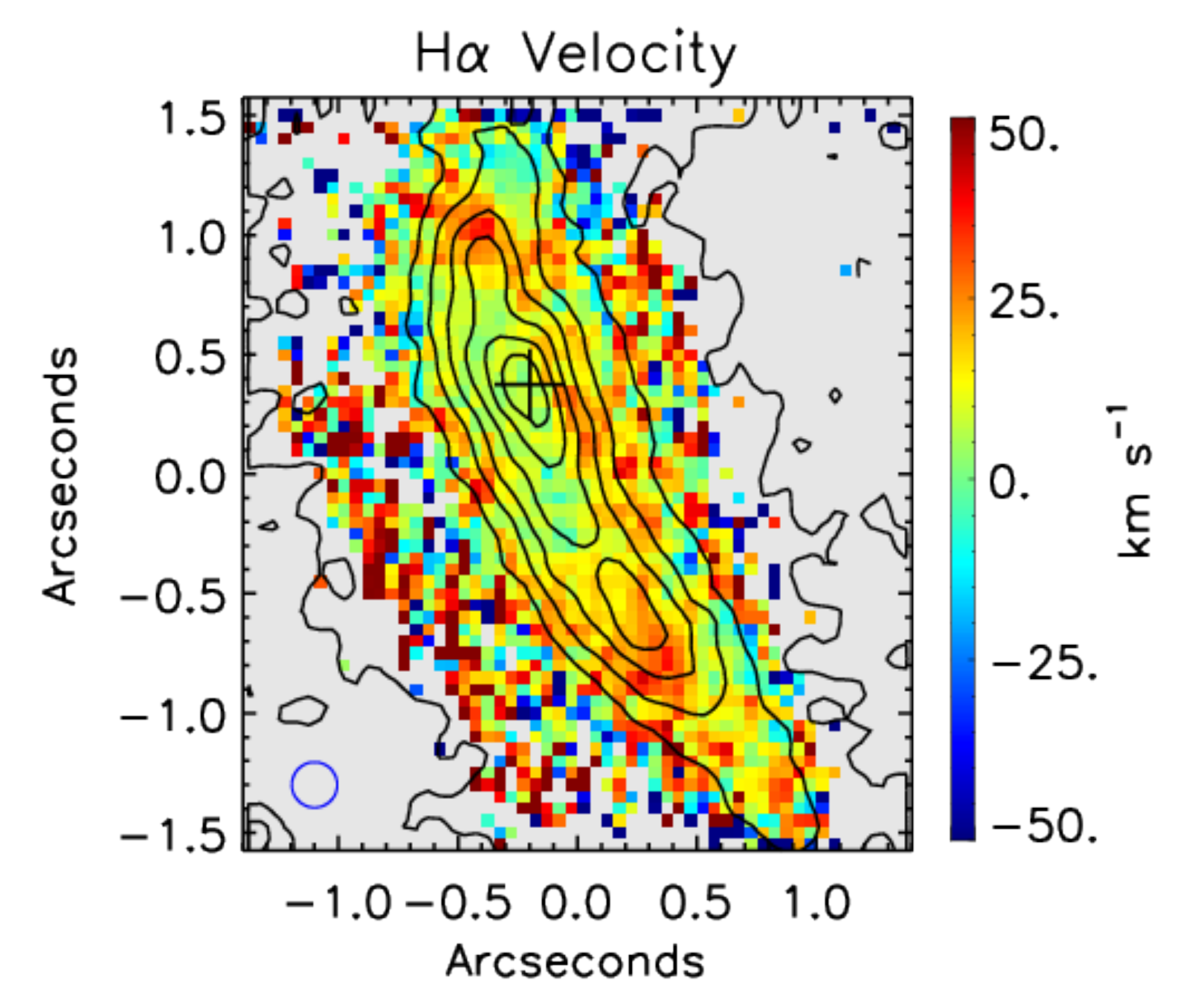}
\includegraphics[width=0.31\textwidth]{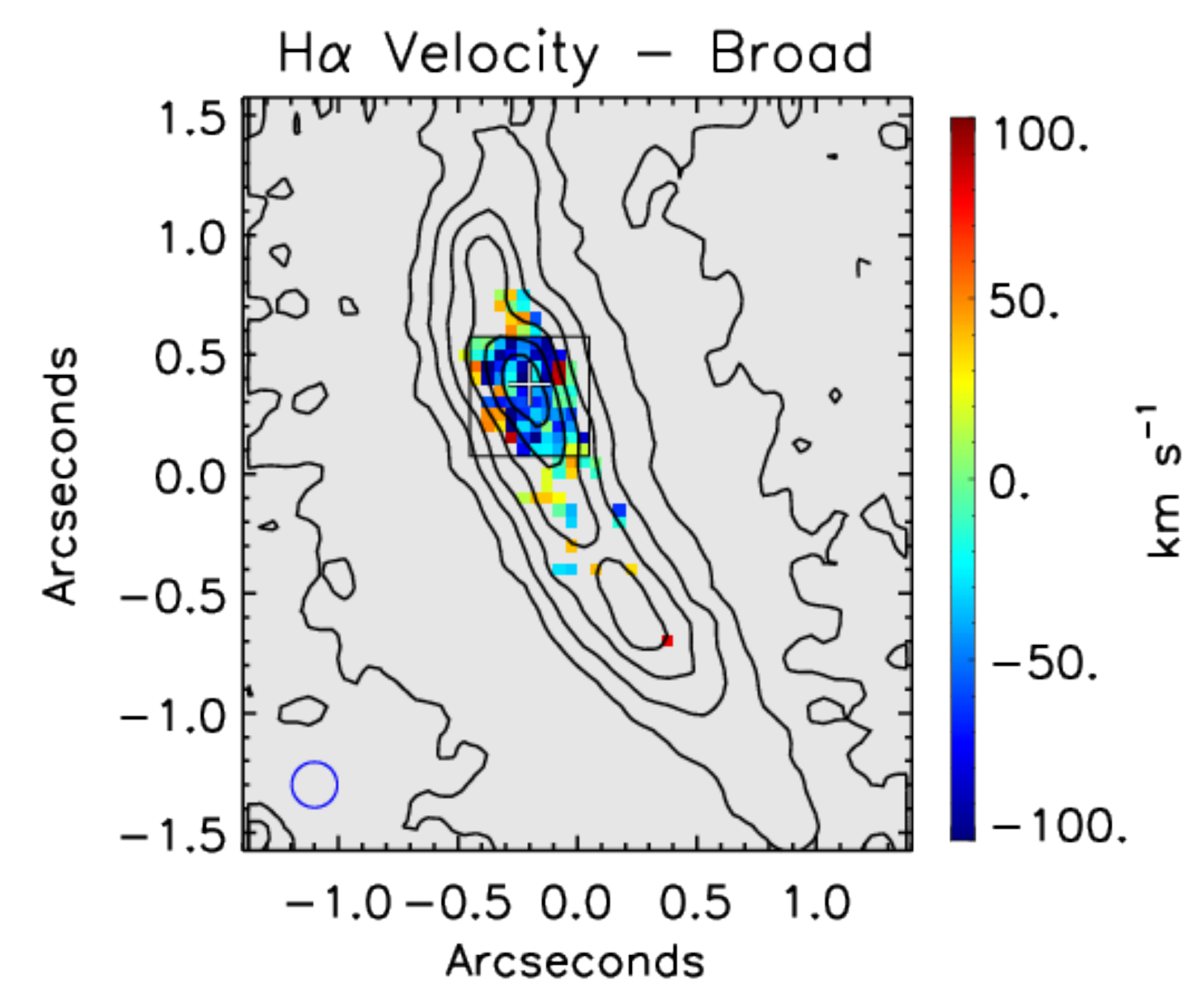}
\includegraphics[width=0.31\textwidth]{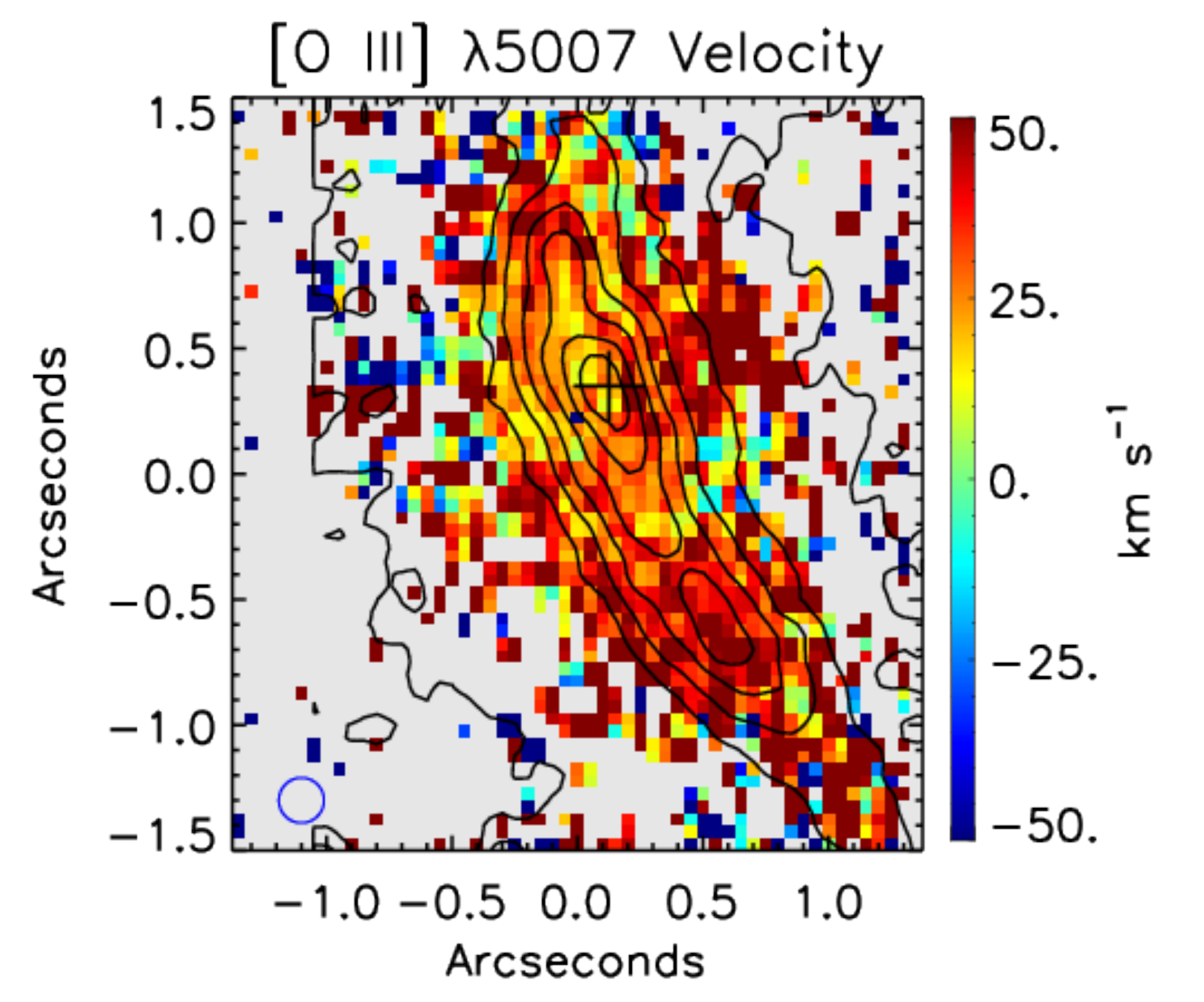}\\
\includegraphics[width=0.31\textwidth]{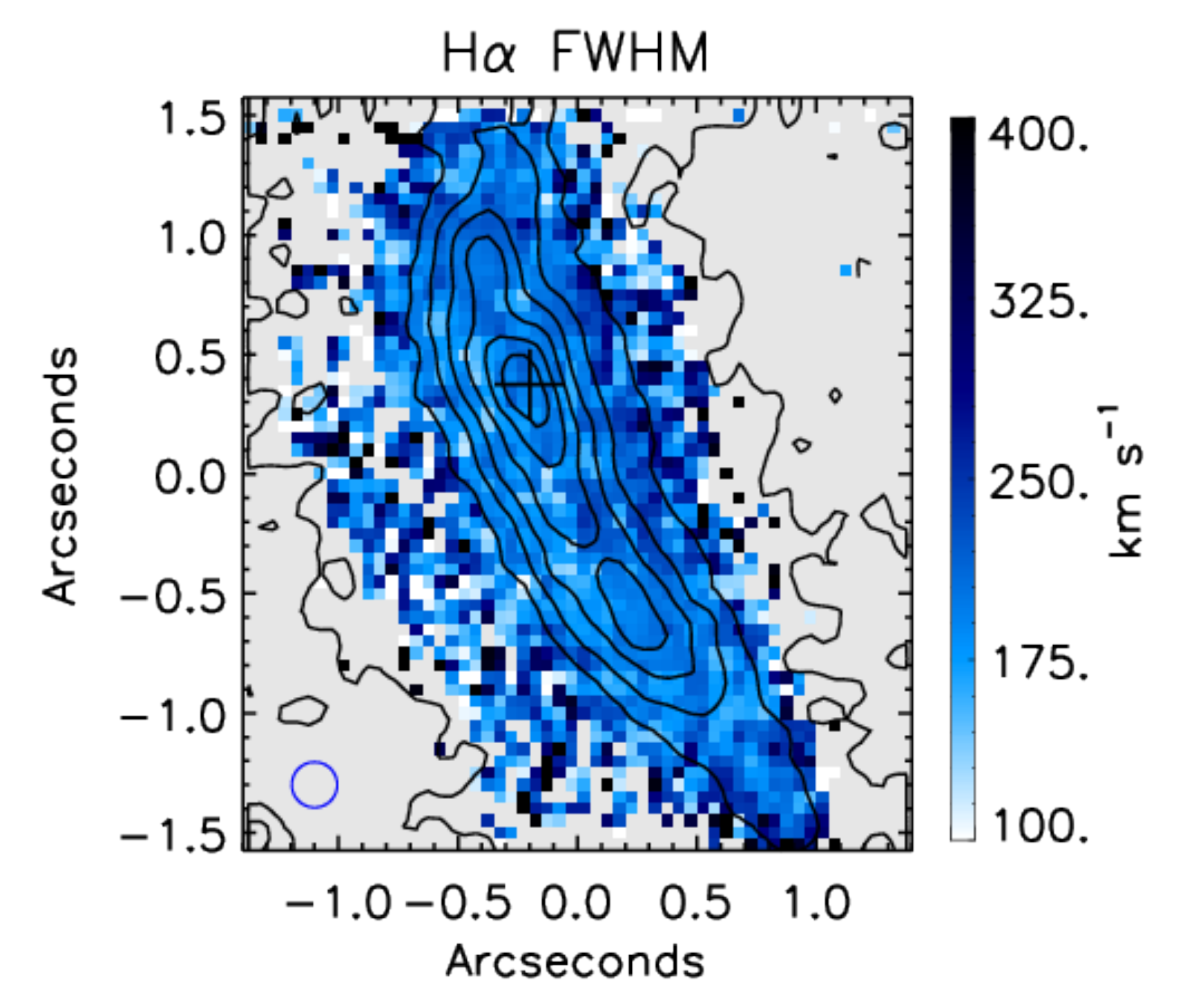}
\includegraphics[width=0.31\textwidth]{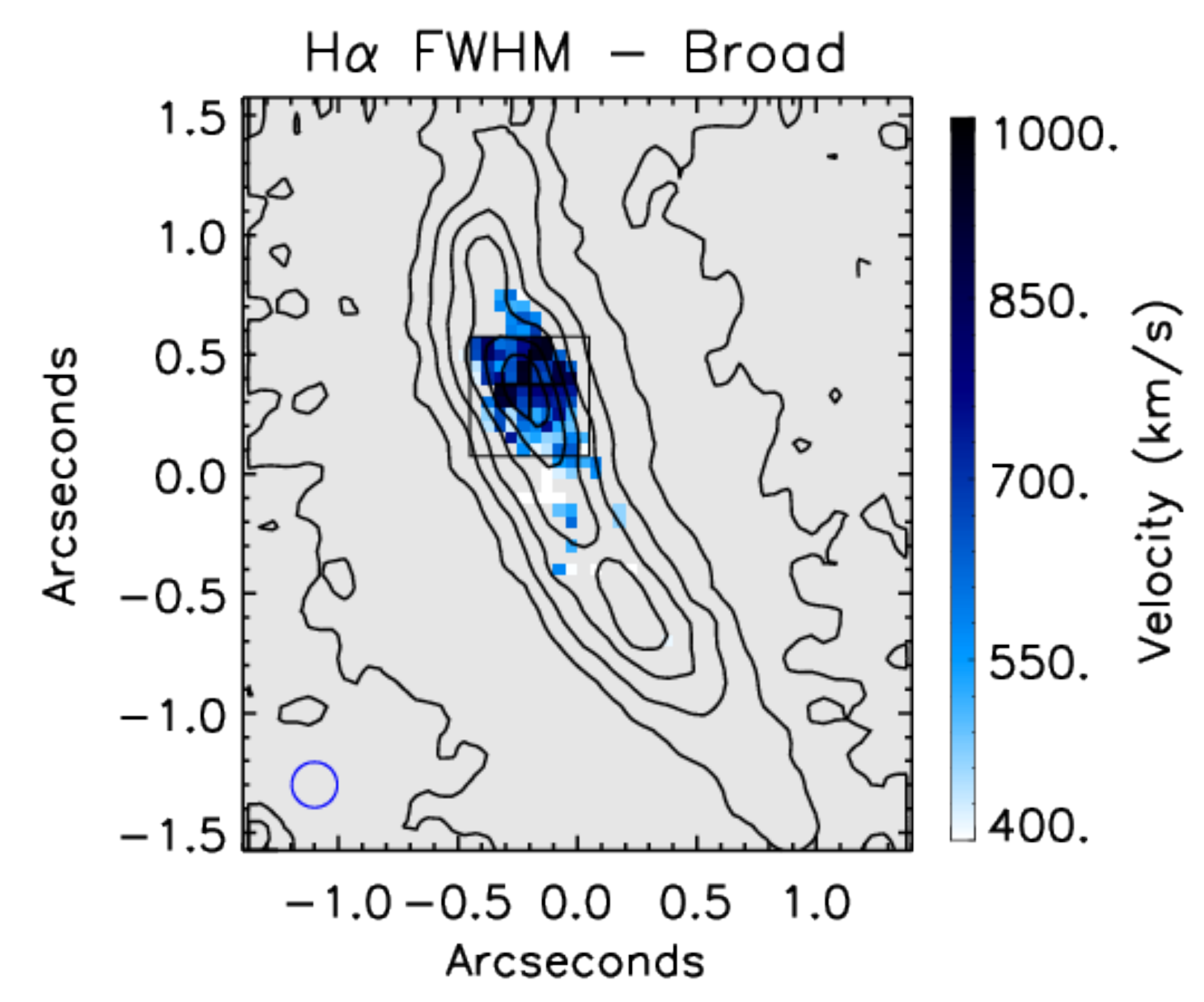}
\includegraphics[width=0.31\textwidth]{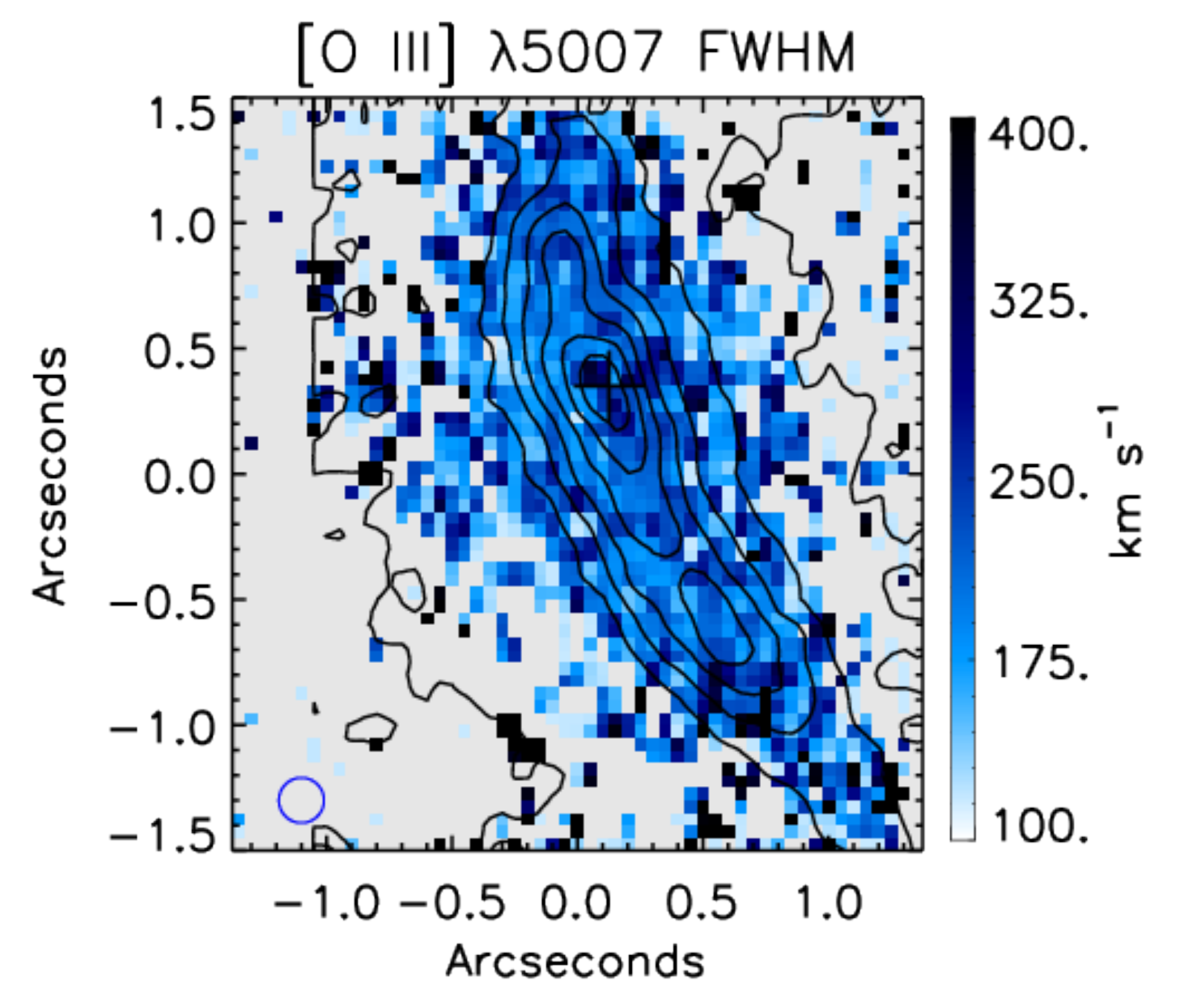}\\
\includegraphics[width=0.31\textwidth]{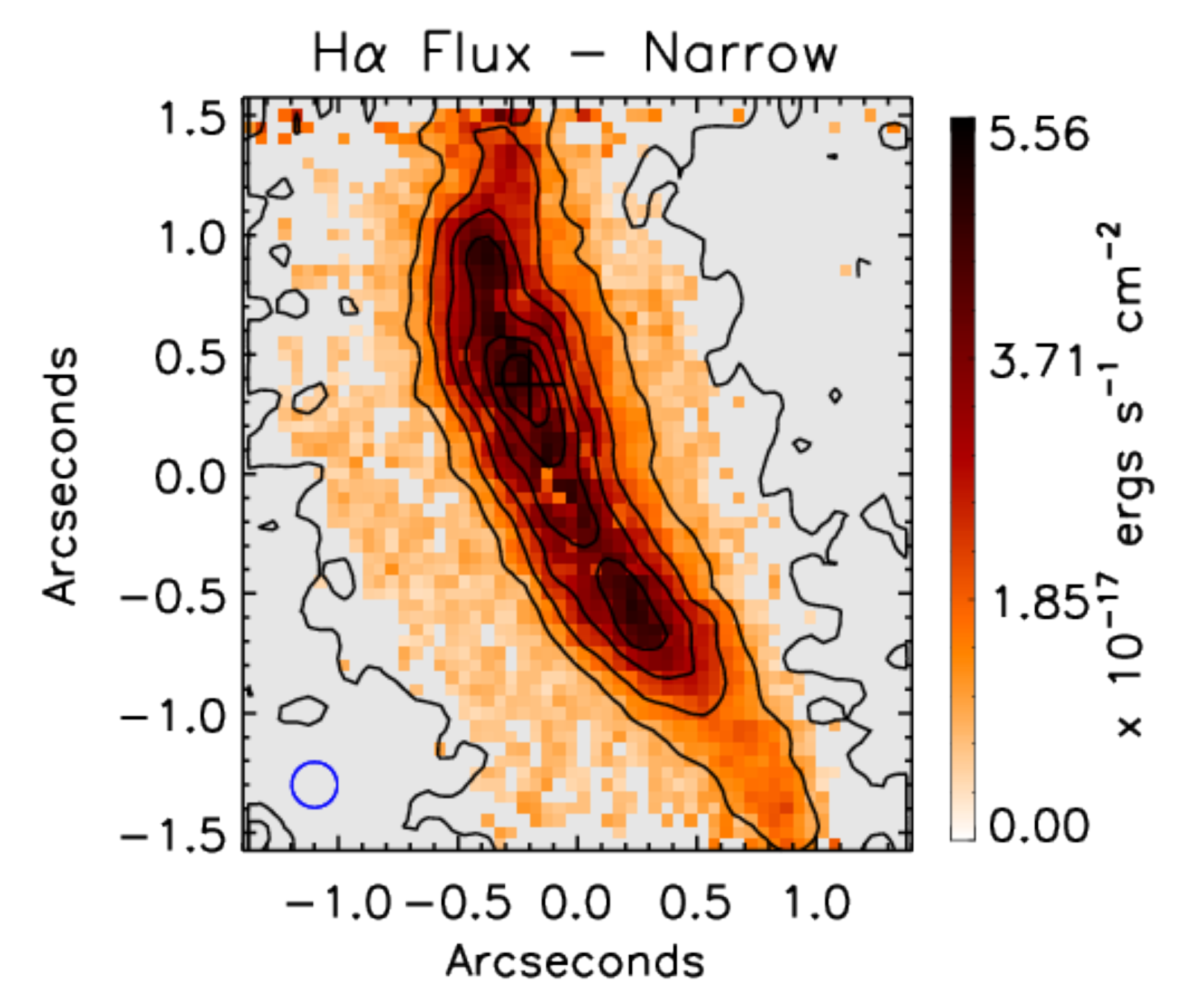}
\includegraphics[width=0.31\textwidth]{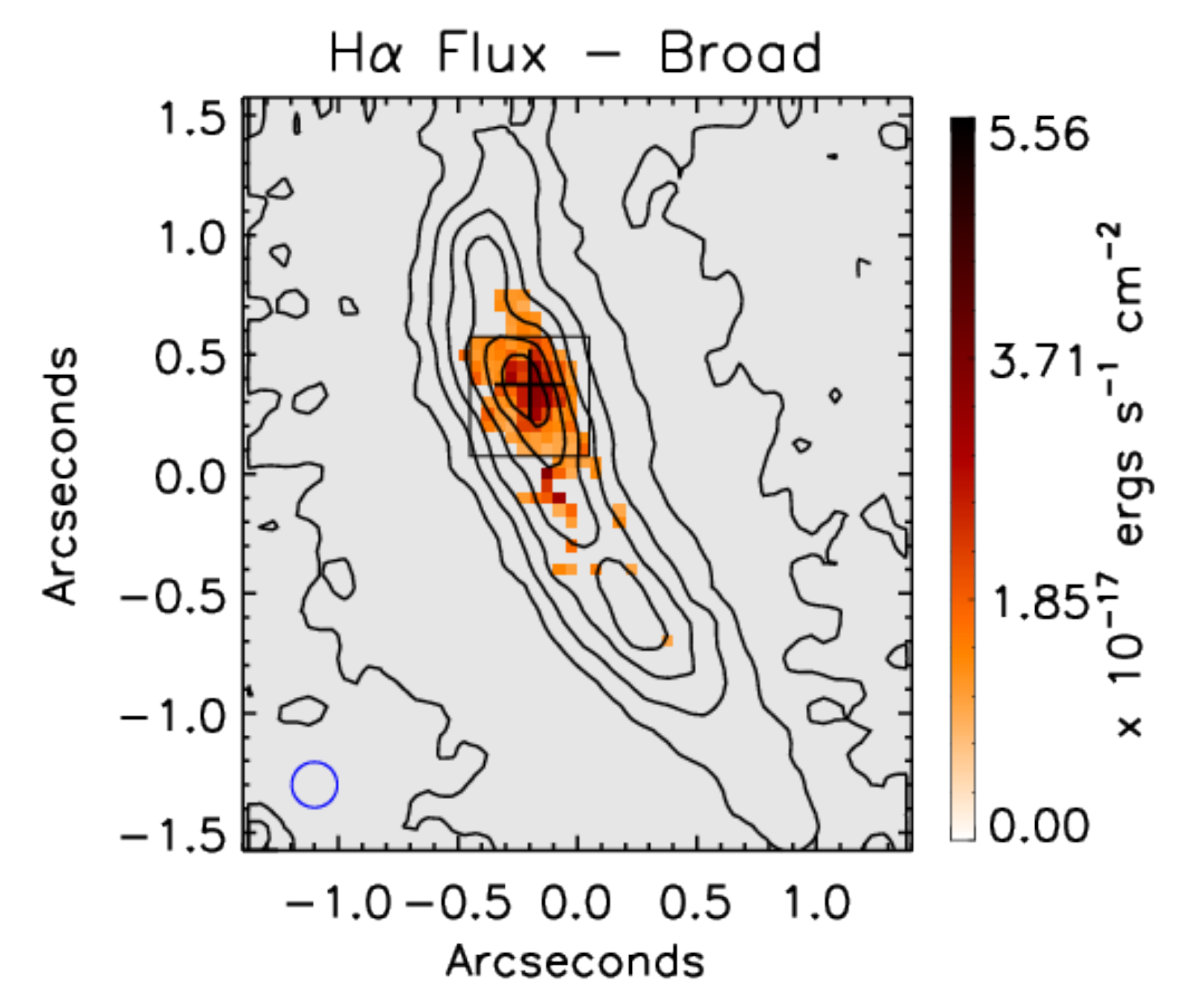}
\includegraphics[width=0.31\textwidth]{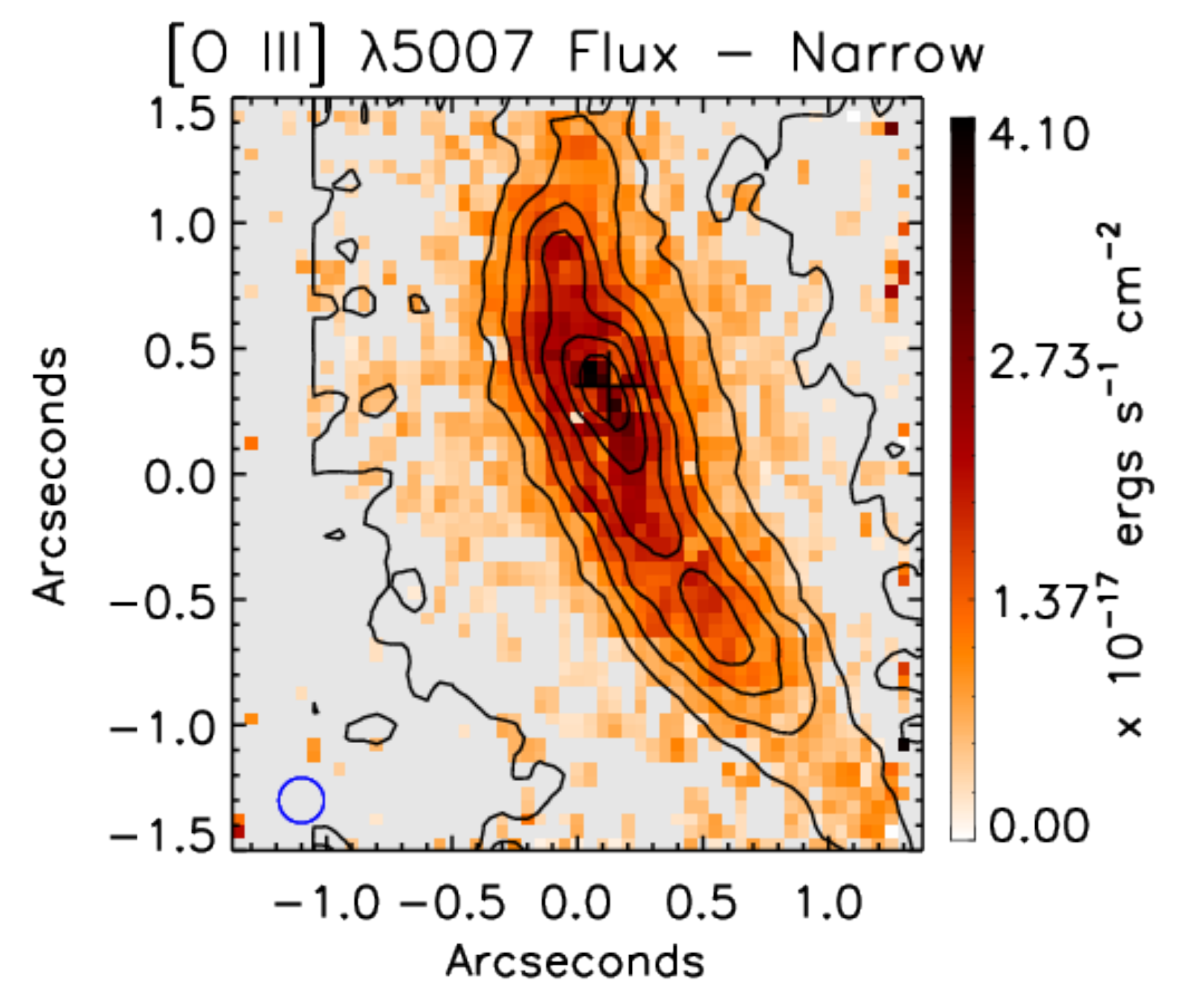}\\
\includegraphics[width=0.31\textwidth]{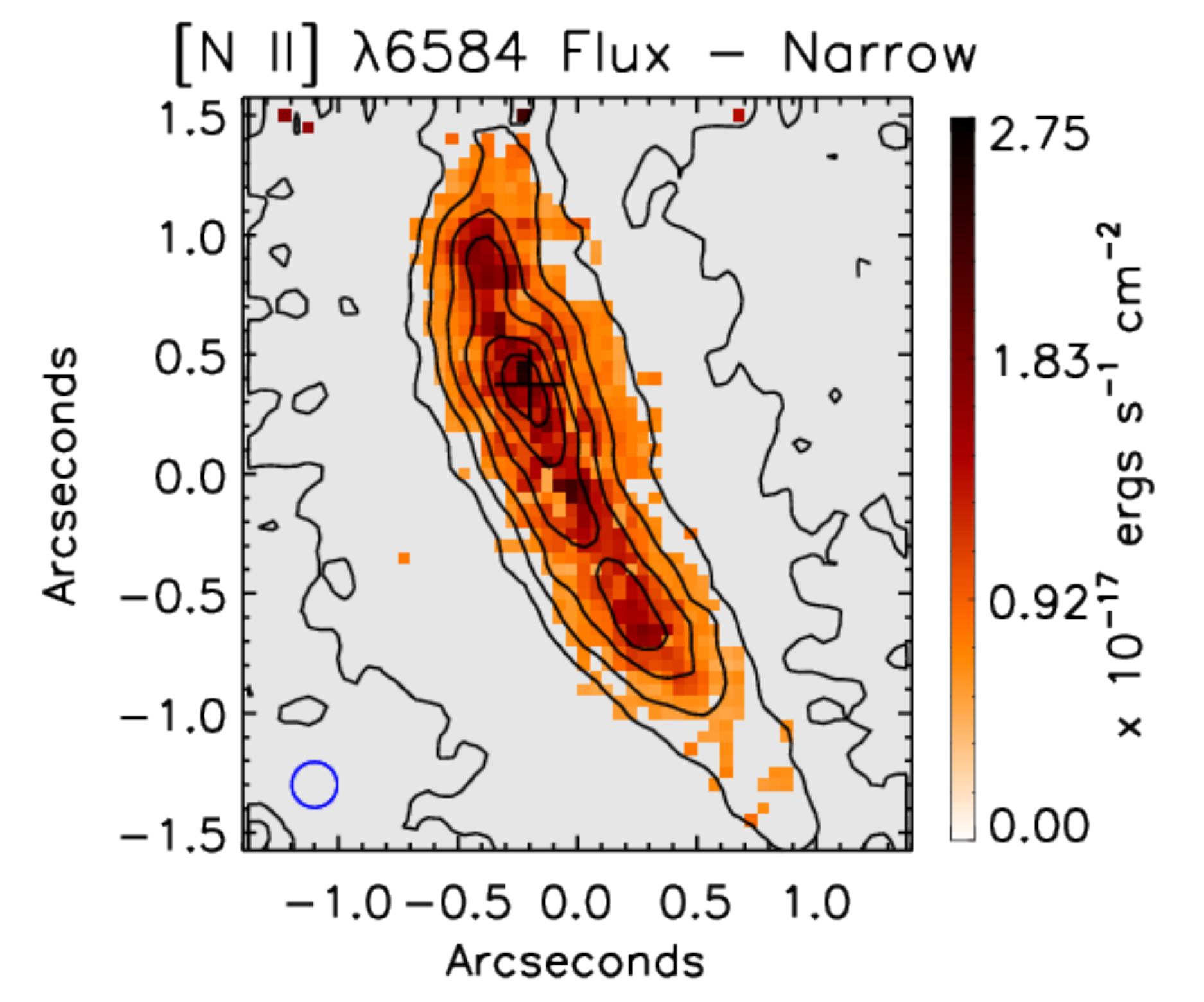}
\includegraphics[width=0.31\textwidth]{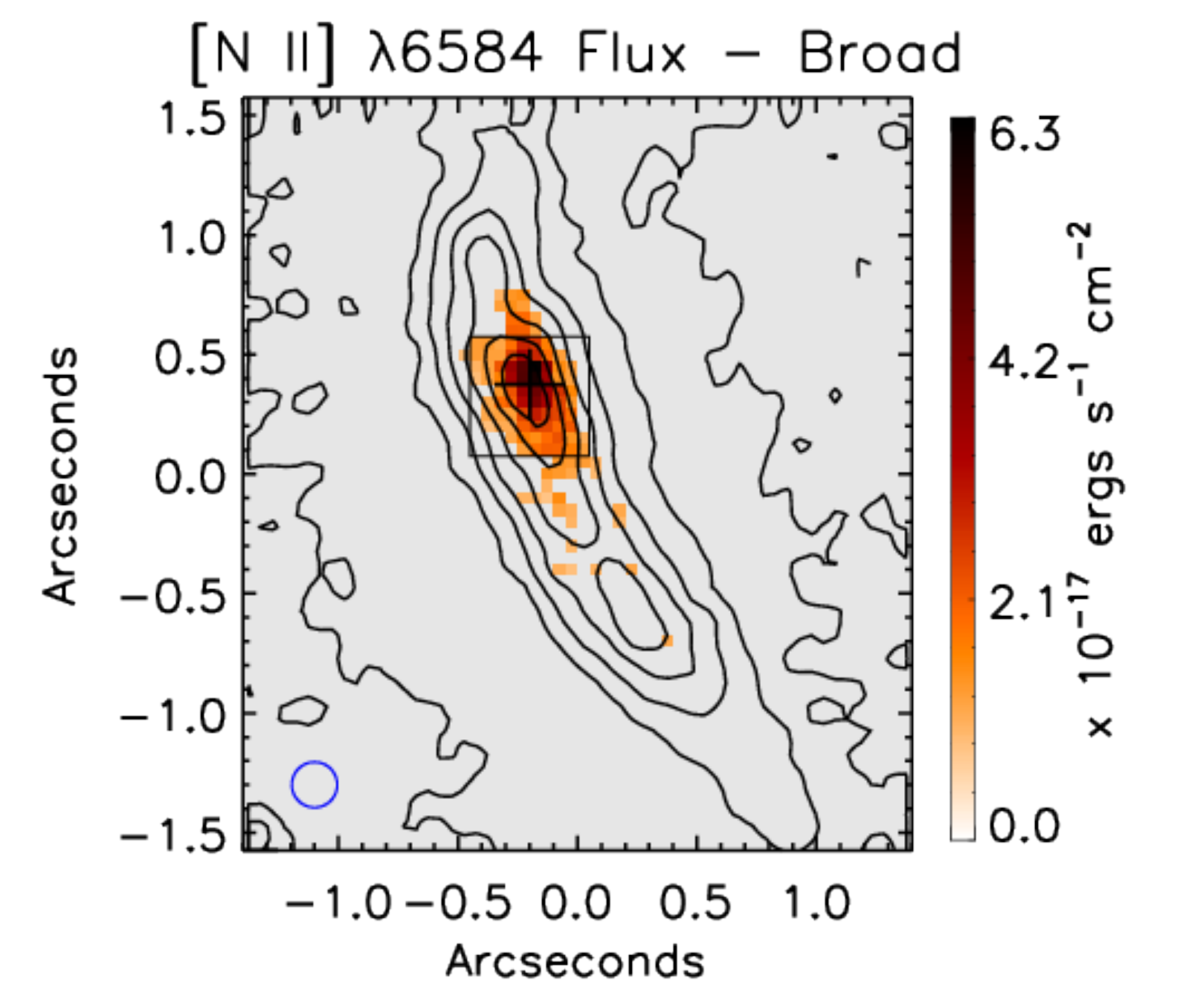}
\includegraphics[width=0.31\textwidth]{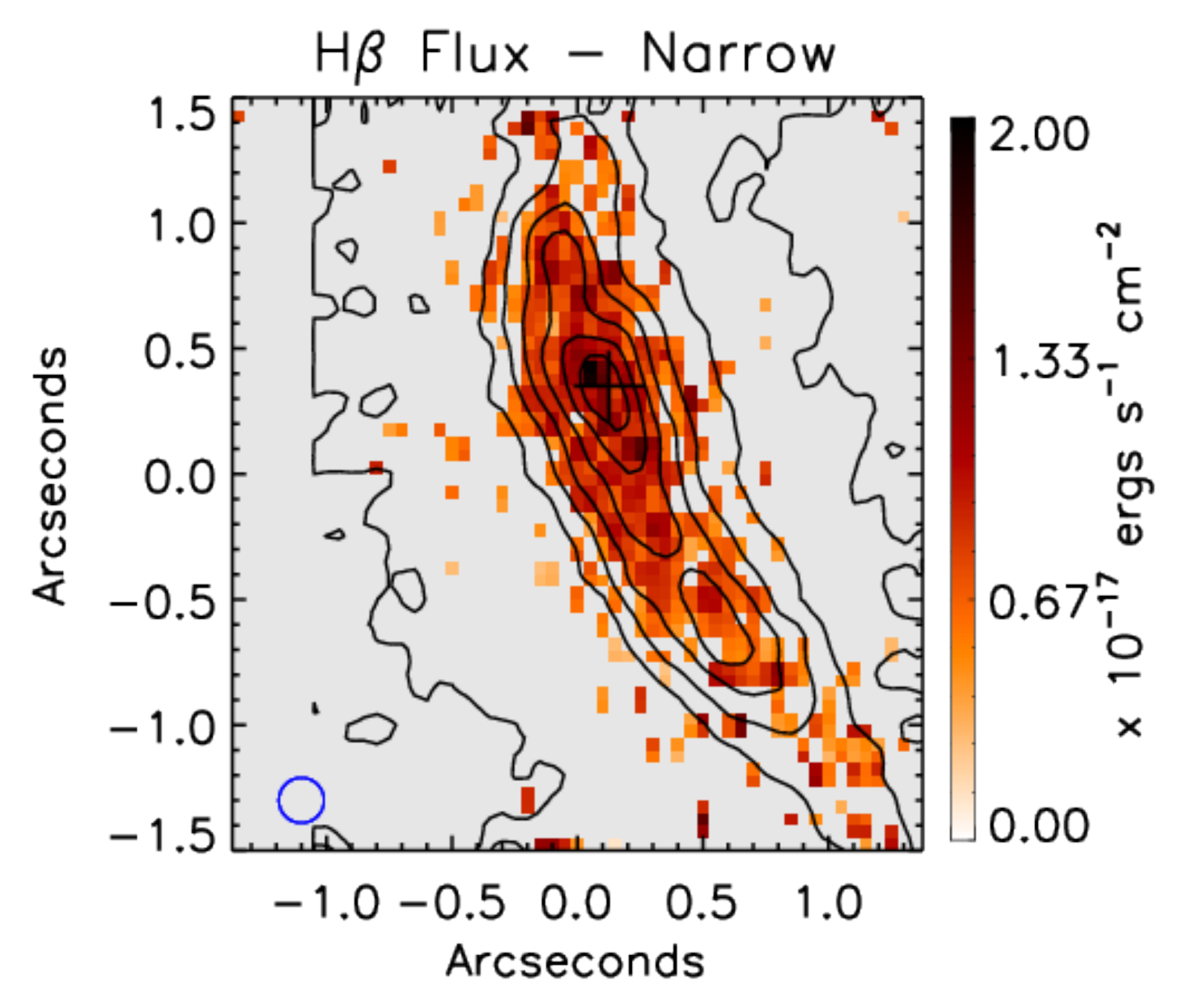} 

\caption{SGAS 0033$+$2 emission line measurements from VLT/SINFONI H- and K-band IFU observations. First, second, and third rows 
display centroid velocity, FWHM, and integrated flux maps, respectively, of H$\alpha$ and [O~III] $\lambda$5007 
emission-line profiles in the image plane. Fourth row displays integrated flux maps of [N~II] $\lambda$6584 and 
H$\beta$. H$\alpha$ and [N~II] emission-line profiles are separated into narrow- and broad-components in the left 
and central columns, respectively. The black boxes in the center column depict a 0.5$" \times$ 0.5$"$ region over the 
central outflowing gas, binned to detect broad-component signatures of [O~III] and H$\beta$. Black contours represent 
integrated, continuum-subtracted H$\alpha$ flux images. The K-band (rest frame optical) continuum flux peak is depicted 
by a cross.}
\label{fig:hamaps}

\end{figure*}

\section{Image Plane Analysis}

\subsection{SINFONI Spectroscopic Fitting}

Emission-line kinematics and fluxes of H$\alpha$, [N II], [O~III], and H$\beta$ were measured in each spaxel of our 
SINFONI H- and K-band data cubes by fitting Gaussians in an automated routine. 
Our fitting process, previously discussed in depth in \citet{Fis17}, uses the Importance Nested Sampling algorithm 
as implemented in the MultiNest library \citep{Fer08,Fer09,Fer13,Buc14} to compute the logarithm of the evidence, 
$lnZ$, for models containing a continuum plus zero to three Gaussian components per emission line. Gaussians were defined 
using Gaussian parameters $\mu$ (centroid), $\sigma$ (dispersion), and H (peak height). When 
comparing two models, i.e. a model with zero Gaussians ($M_{0}$) and a model with one Gaussian ($M_{1}$), the simpler 
model is chosen unless the more complex model, $M_{1}$, has a significantly better evidence value, 
$|ln(Z_{1}/Z_{0})| > 5$ (99\% more likely). Fits of emission lines in individual spaxels used different models for 
each band. H-band models first measured [O~III] $\lambda$5007, simultaneously fitting a second set of components 
to [O~III] $\lambda$4959 in order to properly account for flux contributions from wing emission between both lines, 
and then tested for the presence of H$\beta$. Gaussian wavelength centroid and disperson parameters of [O~III] 
$\lambda$4959 components were fixed following parameters used in fitting [O~III] $\lambda$5007 components, 
with the flux of [O~III] $\lambda$4959 fixed to be 1/3 that of the [O~III] $\lambda$5007 flux. Gaussian wavelength centroid and 
disperson parameters of H$\beta$ components were fixed in the same manner, as we assume that the lines originate from 
the same emission region, and the H$\beta$ flux was left as an open parameter. K-band models first measured H$\alpha$ and then tested for the 
presence of [N II] $\lambda\lambda$6548,6584. Gaussian wavelength centroid and disperson parameters of [N II] 
$\lambda\lambda$6548,6584 were also fixed following parameters used in fitting H$\alpha$, again under the 
assumption that the lines originate from the same emission region, with the flux of [N II] $\lambda$6548 fixed 
to be 1/3 that of the [N II] $\lambda$6584 flux, which was left as an open parameter. 

Initial input parameters in our models are selected based on physical considerations. The centroid position 
for each Gaussian was limited to a 40\AA~range around the wavelength that contained the entirety of the line 
profiles throughout each data cube. Gaussian standard deviation ranged from the spectral resolution of the 
H- and K-band gratings, to an artificial FWHM limit of $\sim$800\,km s$^{-1}$. Gaussian height was defined to 
allow for an integrated flux that ranged from a 3$\sigma$ detection to a maximum integrated flux of 3$\sigma 
\times$ 10$^4$.

Fits from the H- and K-band observations are mapped in Figure \ref{fig:hamaps}. 
Observed velocity, FWHM, and integrated fluxes are shown for H$\alpha$ and [O~III]$\lambda$5007, with 
additional integrated fluxes for H$\beta$ and [N II], as their velocity and FWHM measurements are identical 
to [O~III] and H$\alpha$, respectively. Doppler-shifted velocities are given in the rest frame of the galaxy 
using air rest wavelengths of each line. We found emission lines present in most spaxels to be 
best fit with a single Gaussian, with H$\alpha$ and [N II] emission lines containing two-component line 
profiles in spaxels surrounding the K-band continuum peak (shown as a cross in each map of Figure \ref{fig:hamaps}) 
of the lensed galaxy arc. Two-component fits are sorted by FWHM into separate H$\alpha$ / [N II] maps in Figure 
\ref{fig:hamaps}. Component blending due to lower signal-to-noise ratios for the broad component is observed 
in regions between fits with different numbers of components, as a jump in line dispersion is observed in the narrow-component 
FWHM plot at the border between single and double component fits. 

We find a majority of the emission-line gas fit with single-components, or the narrower of two components, 
is near systemic velocity or slightly redshifted. Emission-line knots north and south of the continuum peak 
show symmetric redshifted kinematics. Additional faint filaments observed in H$\alpha$ and [O~III] east and west of the continuum 
peak also show symmetric redshifted velocities. The broad H$\alpha$ and [N II] emission-line components 
over the continuum peak are typically blueshifted, with an average FWHM of $\sim$540 km/s and maximum 
and average offsets of $\sim$-140\,km s$^{-1}$ and $\sim$-40\,km s$^{-1}$, respectively. We measure the 
spatially-resolved maximum extent of the broad-FWHM, blueshifted gas in the image plane by fitting the region with 
an ellipse of r$_{maj}$= 0.35$''$, b/a = 0.4, and PA = 30$\degree$ east of north. At 7" from the guide star, we 
note a degradation of the reported K-band PSF is expected, with the Strehl Ratio of observations for SGAS 0033+2 
decreasing by approximately 20$\%$ per the SINFONI User Manual. Temporal variations of the atmosphere also add 
uncertainty on the effective PSF during the observations, with individual exposures of PSF stars in 
similar observations by \cite{For18} indicating typical OB-to-OB variations of $\sim$30$\%$ in PSF FWHM. As such, assuming 
an effective PSF during the observations to be $\sim$ 0.3$''$, the spatial extent of the observed outflows 
remains well resolved.

\begin{figure*}[h]
\centering
\includegraphics[width=0.48\textwidth]{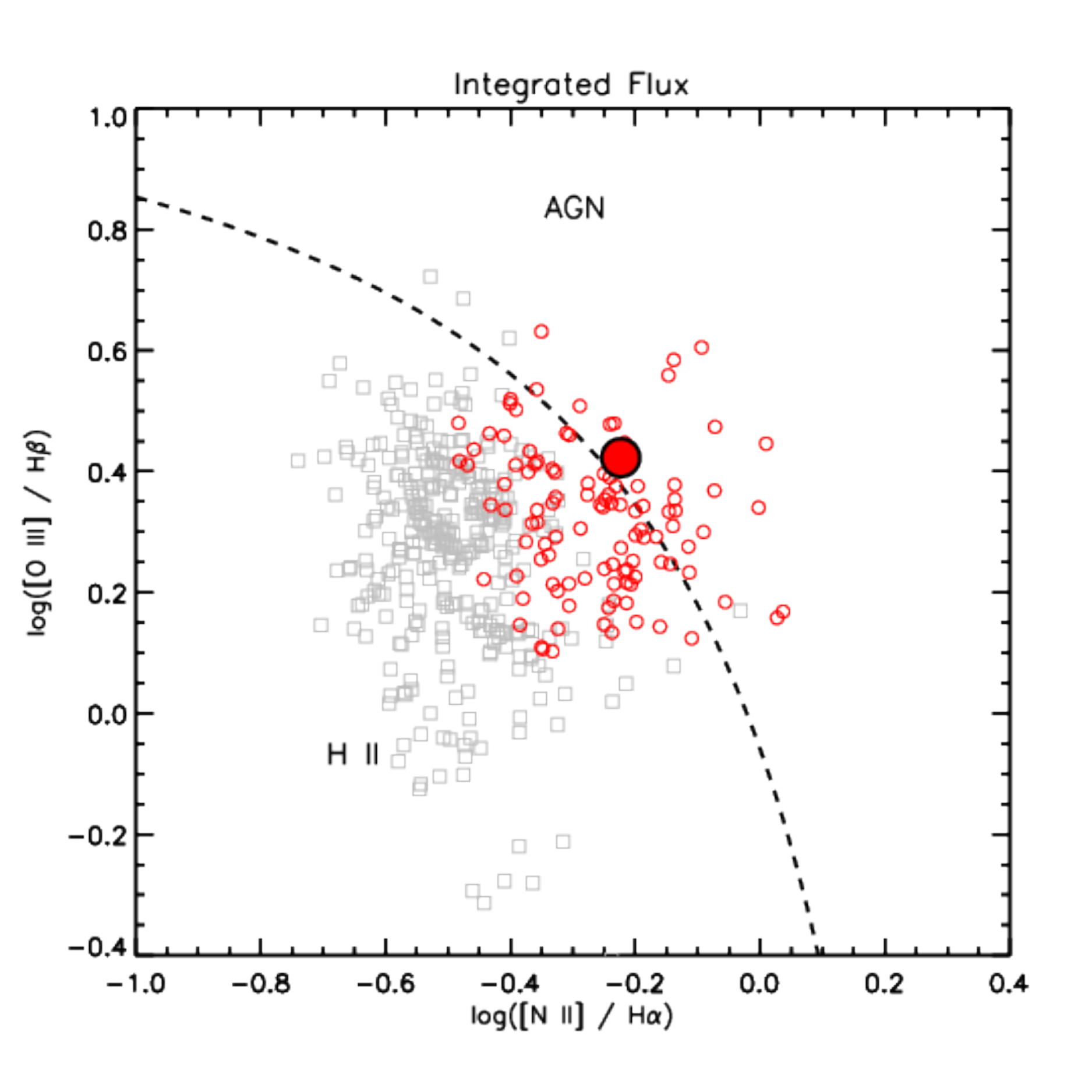}
\includegraphics[width=0.48\textwidth]{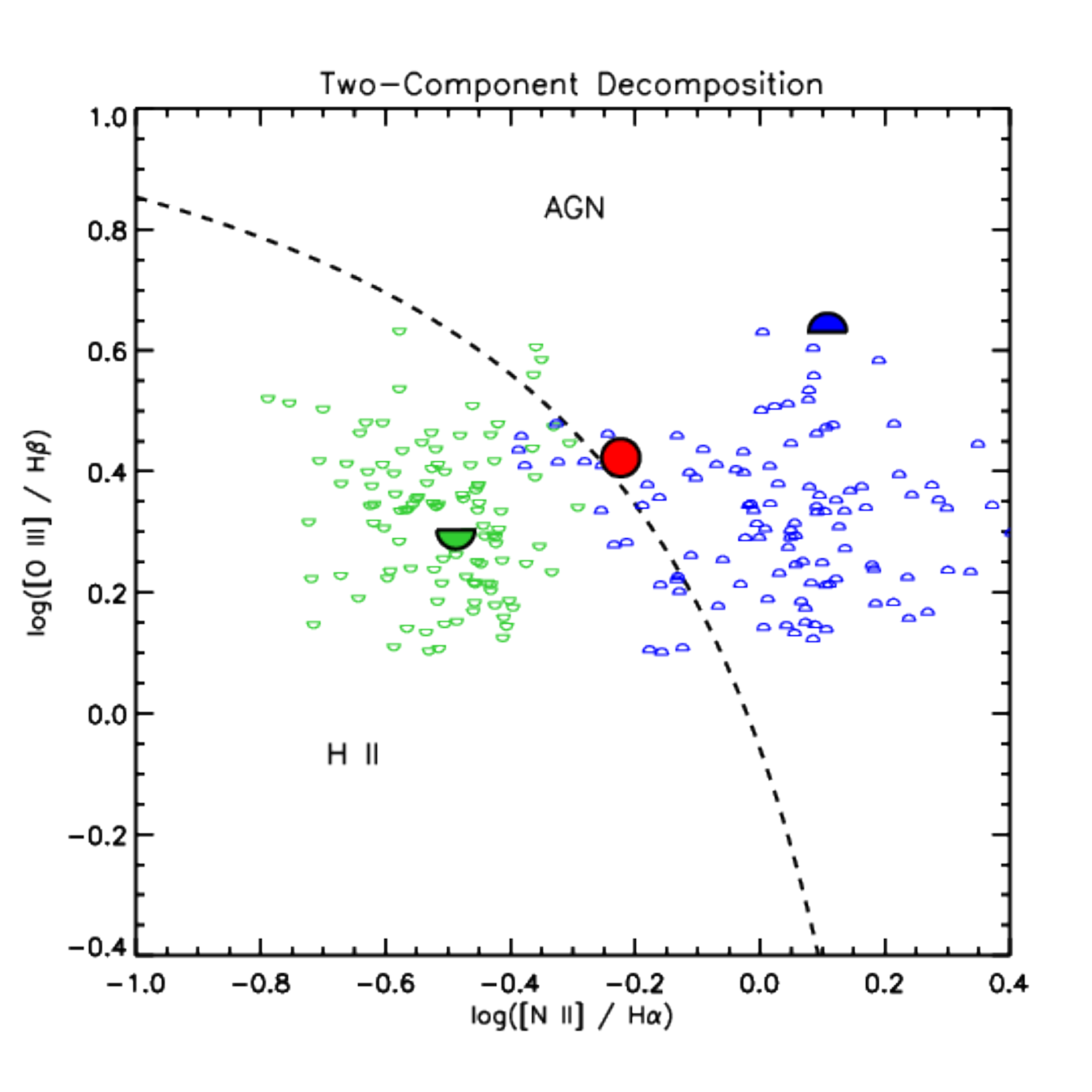}

\caption{[N II]/H$\alpha$ vs [O~III]/H$\beta$ diagnostic diagrams derived from the H- and K-band SINFONI 
observations in the image plane. Left: Diagnostic diagram for individual spaxel measurements using the total, 
integrated flux of each emission-line. The division between H II and AGN ionization is defined by the redshift-dependent 
classification from \citealt{Kew13}. Grey squares represent ratios from spaxels with fluxes measured using a single 
line component in all measured lines. Red open circles represent ratios from spaxels fluxes measured using a single 
line component in [O~III] and H$\beta$ and two components for [N II] and H$\alpha$. The red filled circle shows the 
ratio measured from a binned spectrum containing all spaxels with two emission-line components. Right: Diagnostic 
diagram for individual spaxel measurements with the two component [N II] and H$\alpha$ 
emission-lines decomposed into narrow and broad components. Green open lower-half circles represent narrow-component 
fluxes and blue open upper-half circles represent broad-component fluxes. The red filled circle again shows the 
ratio measured from a binned spectrum containing all spaxels with two emission-line components, and the filled green 
lower-half circles and blue open upper-half circles show the ratio measured from the narrow- and broad-components of 
the binned spectrum, respectively. }

\label{fig:bpt}

\end{figure*}

\begin{figure*}[h]
\centering
\includegraphics[width=0.98\textwidth]{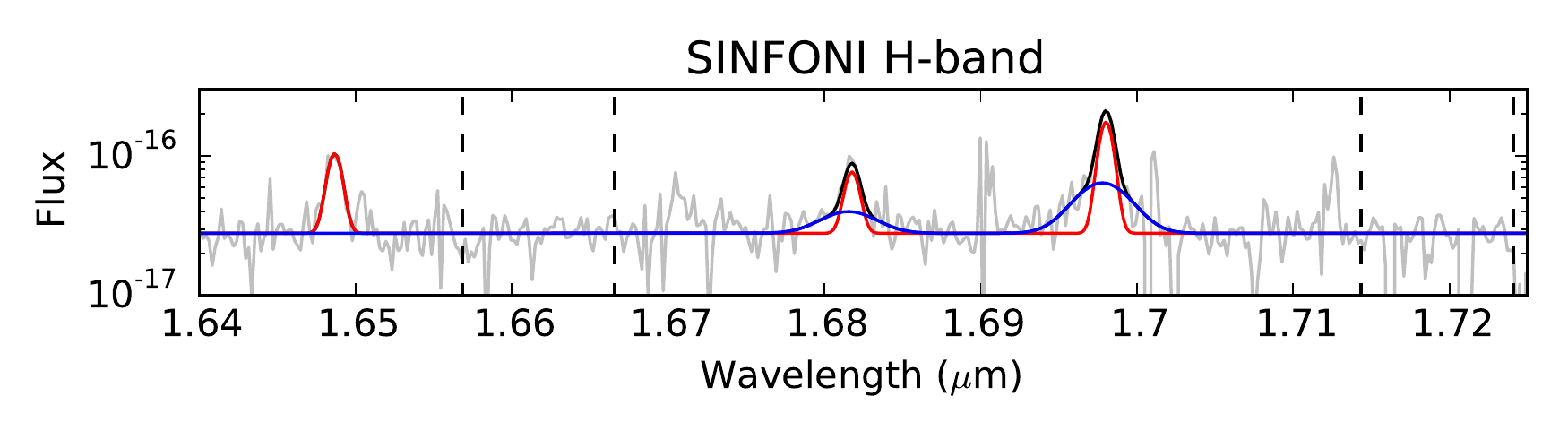}\\
\vspace{-.5cm}
\includegraphics[width=0.98\textwidth]{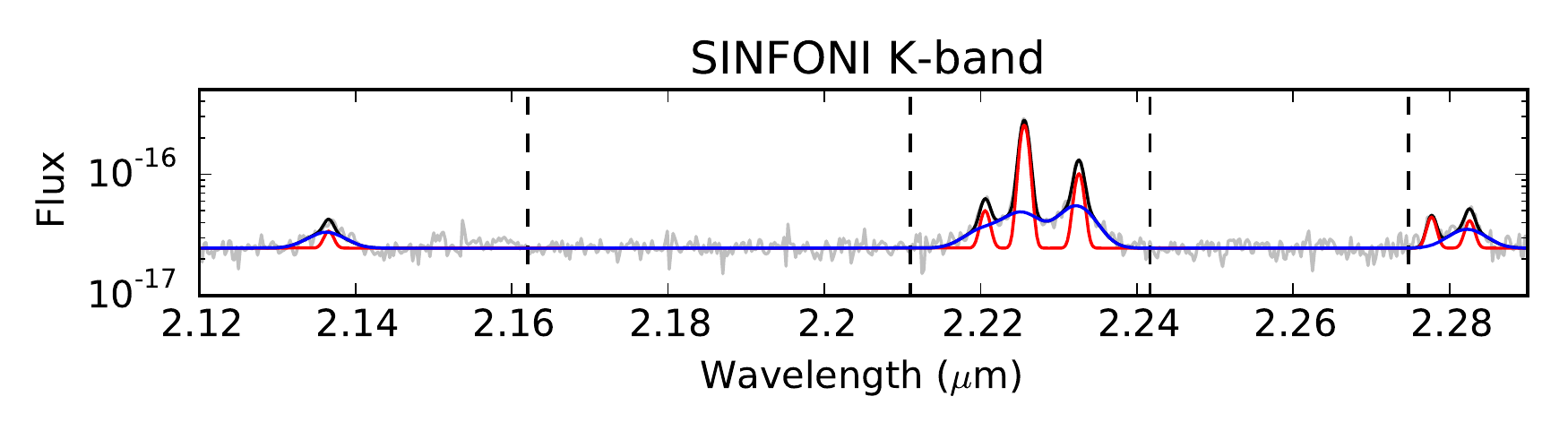}

\caption{Gaussian fits to the binned SINFONI spectrum of the central 0.5$" \times$ 0.5$"$ region 
over the lensed arc image plane continuum peak. Top: H-band spectrum fitting H$\beta$ and [O~III]$\lambda\lambda$4959,5007. 
Bottom: K-band spectrum fitting [O I] $\lambda$ 6300, H$\alpha$, [N II]$\lambda\lambda$6548,6584, and [S II]$\lambda\lambda$6716,6731. 
Gaussian fits to [O~III] $\lambda$4959 and [N II]$\lambda$6548 used height parameters fixed to be 1/3 of 
[O~III]$\lambda$5007 and [N II]$\lambda$6584, respectively. No broad-component is detected for H$\beta$ 
and [S II]$\lambda$6716. Grey line represents SINFONI spectral data. Solid black line represents the 
total model. Blue and red lines represent decompositions of broad and narrow Gaussian components, respectively. 
Flux is in units of erg s$^{-1}$ cm$^{-2}$ \AA$^{-1}$.}
\label{fig:binlinefit}

\end{figure*}

\begin{table*}[h]
  \centering
  \ra{1.3}
  \caption{Central 0.5$" \times$ 0.5$"$ Binned AGN Flux Measurements}
  \label{tab:binflux}
  \tabcolsep=0.1cm
  \begin{tabular}{lclll}
    \toprule	
    Line					    &       	   & Observed                           & Observed                  & Dereddened    \\
                                & FWHM         & Image Plane                & Source Plane              & Source Plane  \\
  							    & (km s$^{-1}$)& (erg s$^{-1}$ cm$^{-2}$)   & (erg s$^{-1}$ cm$^{-2}$)  &(erg s$^{-1}$ cm$^{-2}$)  \\
  	\midrule
  	Narrow Component & & & & \\

    H$\beta$ 	                & 190          & 8.77\,$\pm$\,2.68$\times$10$^{-16}$	& 2.80\,$\pm$\,1.04$\times$10$^{-18}$	& 6.94\,$\pm$\,2.58$\times$10$^{-18}$    \\  	
    $[$O III$] \lambda$5007     & 190          & 1.70\,$\pm$\,0.49$\times$10$^{-15}$   	& 5.56\,$\pm$\,2.45$\times$10$^{-18}$   & 1.34\,$\pm$\,0.59$\times$10$^{-17}$    \\ 
    $[$O I$] \lambda$6300	    & 190          & 1.42\,$\pm$\,0.73$\times$10$^{-16}$	& 4.52\,$\pm$\,2.87$\times$10$^{-19}$	& 8.68\,$\pm$\,5.51$\times$10$^{-19}$    \\
    H$\alpha$                   & 190          & 3.49\,$\pm$\,0.48$\times$10$^{-15}$	& 1.07\,$\pm$\,0.15$\times$10$^{-17}$	& 1.98\,$\pm$\,0.29$\times$10$^{-17}$    \\
    $[$N II$] \lambda$6584		& 190          & 1.16\,$\pm$\,0.16$\times$10$^{-15}$    & 3.46\,$\pm$\,0.50$\times$10$^{-18}$   & 6.39\,$\pm$\,0.92$\times$10$^{-18}$    \\
    $[$S II$] \lambda$6716	    & 190          & 2.98\,$\pm$\,1.06$\times$10$^{-16}$	& 8.43\,$\pm$\,3.49$\times$10$^{-19}$	& 1.53\,$\pm$\,0.63$\times$10$^{-18}$	 \\
    $[$S II$] \lambda$6731	    & 190          & 2.48\,$\pm$\,0.86$\times$10$^{-16}$ 	& 7.65\,$\pm$\,3.17$\times$10$^{-19}$ 	& 1.39\,$\pm$\,0.58$\times$10$^{-18}$	\\
    \midrule    
    Broad Component & & & & \\
    
    H$\beta$ 	                & 705		    & 3.34\,$\pm$\,1.02$\times$10$^{-16 \rm ~a}$  & 8.89\,$\pm$\,3.31$\times$10$^{-19 \rm ~a}$  & 2.20\,$\pm$\,0.82$\times$10$^{-18 \rm ~a}$   \\  	
    $[$O III$] \lambda$5007     & 705 		    & 1.16\,$\pm$\,0.34$\times$10$^{-15}$      	     & 3.87\,$\pm$\,1.70$\times$10$^{-18}$      	  & 9.33\,$\pm$\,4.11$\times$10$^{-18}$      		\\ 
    $[$O I$] \lambda$6300	    & 725 		    & 4.89\,$\pm$\,2.52$\times$10$^{-16}$    	     & 1.18\,$\pm$\,0.75$\times$10$^{-18}$    	      & 2.56\,$\pm$\,1.63$\times$10$^{-18}$    			\\
    H$\alpha$                   & 725 		    & 1.33\,$\pm$\,0.18$\times$10$^{-15}$    	     & 3.40\,$\pm$\,0.49$\times$10$^{-18}$    	      & 6.30\,$\pm$\,0.91$\times$10$^{-18}$    			\\
    $[$N II$] \lambda$6584		& 725 		    & 1.72\,$\pm$\,0.24$\times$10$^{-15}$    	     & 4.61\,$\pm$\,0.66$\times$10$^{-18}$    	      & 8.50\,$\pm$\,1.22$\times$10$^{-18}$    			\\
    $[$S II$] \lambda$6716	    & 725 		    & 2.76\,$\pm$\,0.99$\times$10$^{-16 \rm ~b}$  & 8.84\,$\pm$\,3.66$\times$10$^{-19 \rm ~b}$  & 1.61\,$\pm$\,0.67$\times$10$^{-18 \rm ~b}$	\\
    $[$S II$] \lambda$6731	    & 725 		    & 6.40\,$\pm$\,2.28$\times$10$^{-16}$	         & 1.72\,$\pm$\,0.71$\times$10$^{-18}$		      & 3.13\,$\pm$\,1.30$\times$10$^{-18}$				\\
    \bottomrule
  \end{tabular}
  \medskip\\
    \raggedright
    $^{\rm a}$ Estimated assuming H$\alpha_{narrow}$ / H$\beta_{narrow}$ = H$\alpha_{broad}$ / H$\beta_{broad}$ \\
    $^{\rm b}$ 3$\sigma$ detection flux limit \hfill \\[1pt] 

  \vspace{.5cm}
\end{table*}

\subsection{Ionization Source Diagnostics}
We compare measured line flux ratios in an ionization diagnostic diagram (i.e. BPT diagram; 
\citealt{Bal81}) to spatially resolve the source of ionization throughout the image plane arc and 
determine whether the observed blueshifted outflows can be attributed to an AGN. Note that measured 
ratios are not affected by magnification as lensing effects are achromatic. To account for 
the high redshift of our target, we used a redshift-dependent classification that utilizes the standard 
optical diagnostic line ratios [O~III]/H$\beta$ vs [N II]/H$\alpha$ as detailed in \citet{Kew13}. 
Our initial diagnostic diagram, provided in the left plot in Figure \ref{fig:bpt}, compares line 
ratios using the integrated flux across all components of each line. Grey points in this 
distribution have single component fits for each emission line, while red points use summed 
fluxes of H$\alpha$ and [N II]$\lambda$6584 emission lines across both a narrow- and broad-component. 
Decomposing these two component line emission lines into individual narrow- and broad- components to 
obtain their individual ratios, as shown in the right plot of Figure \ref{fig:bpt}, we find that the narrow 
components align with the grey points of the left figure, and that the broad components exhibit an 
[N II]/H$\alpha$ ratio that suggests AGN ionization. Note that the position of the broad 
components on the diagram uses the same [O~III]/H$\beta$ ratio as their corresponding narrow lines 
because broad [O~III]/H$\beta$ components are not observed in individual spaxels.

In order to detect broad-component signatures of [O~III] and H$\beta$, we binned spectra over a 
$0.5" \times 0.5"$ square surrounding the continuum peak and a majority of the blueshifted outflows 
(binned region is shown in the broad-component H$\alpha$ and [N II] maps of Figure \ref{fig:hamaps}). Fits to the 
resultant H- and K-band spectra are shown in Figure \ref{fig:binlinefit}, where we are able to detect 
a broad emission line for [O~III], as well as emission from [O I]$\lambda$6300 and [S II]$\lambda\lambda$ 6716,6731, 
but remain unable to detect broad H$\beta$. Fit parameters for each emission-line in the image-plane binned 
spectra are provided in Table \ref{tab:binflux}. To determine a lower limit on the summed broad-component [O~III]/H$\beta$ 
ratio, we estimate the flux of the unobserved H$\beta$ broad-component to scale to its narrow-component in a similar 
fashion to the observed broad- and narrow-components of H$\alpha$ in the same binned region. The estimated flux 
of 3.34$\times$10$^{-16}$\,erg s$^{-1}$ cm$^{-2}$ is consistent with our measurements, such that the broad H$\beta$ 
would be likely be undetectable compared to the low signal of the brighter [O~III]$\lambda$ 4959 broad-component.

Flux ratios derived from our binned spectrum are plotted as larger, filled points in 
Figure \ref{fig:bpt}, where the red circle, and green and blue half-circles represent flux ratios using 
both components, the narrow component, and the broad component, respectively. We find a lower limit on the broad-component 
[O~III]/H$\beta$ ratio to be 0.54, which is elevated relative to the narrow- and summed-component ratios. 
This suggests that a majority of the broad-components would likely have larger [O~III]/H$\beta$ ratios using their 
true line fluxes instead of estimates and remain in the AGN ionized portion of the diagnostic diagram.
In tandem, the measured emission-line flux ratios and kinematics suggest that we are observing outflows from 
an AGN in the arc image plane of SGAS 0033+02.

\subsection{Spatially Resolved Ly$\alpha$ Structure}

\begin{figure}[h]
\centering
\includegraphics[width=0.48\textwidth]{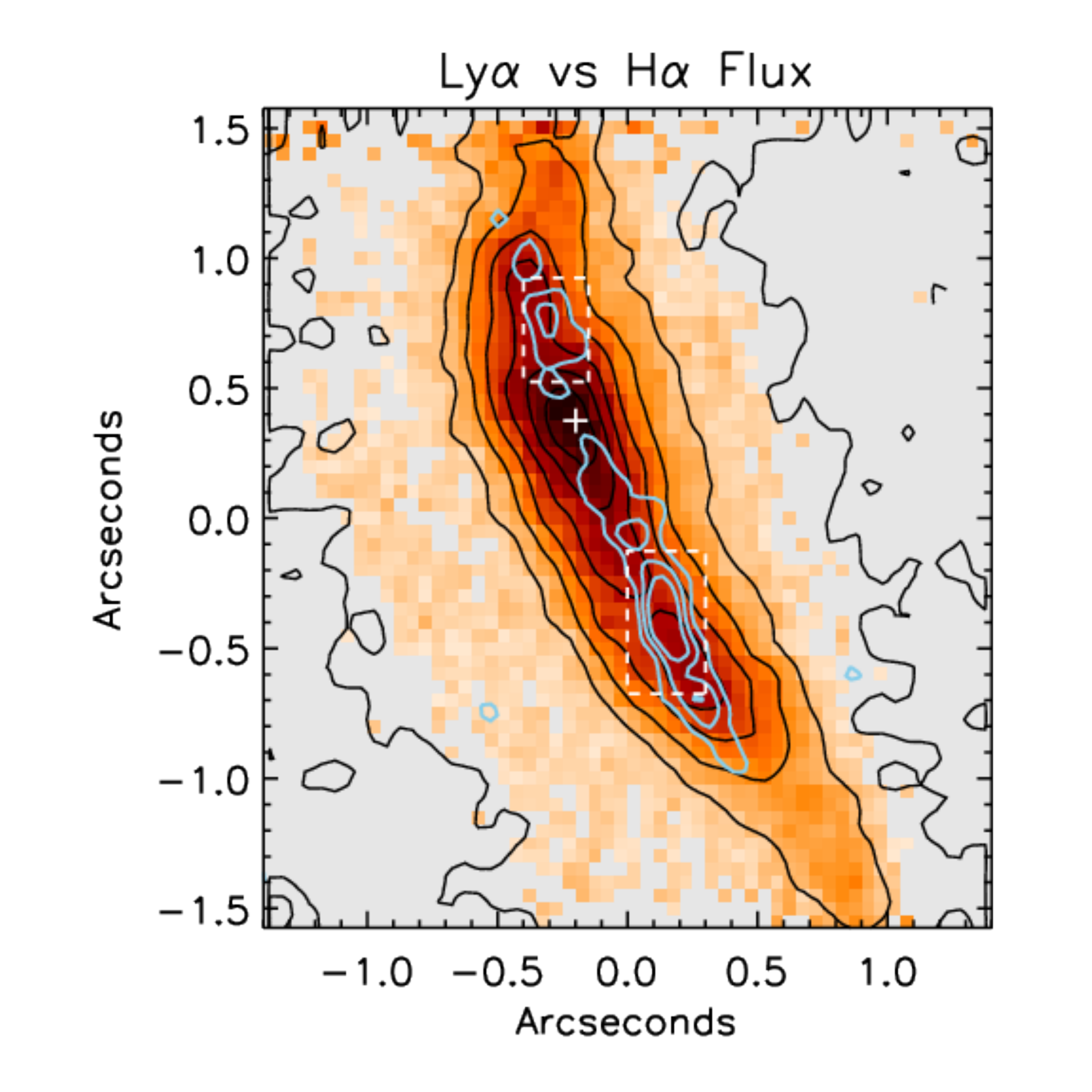}\\
\includegraphics[width=0.4\textwidth]{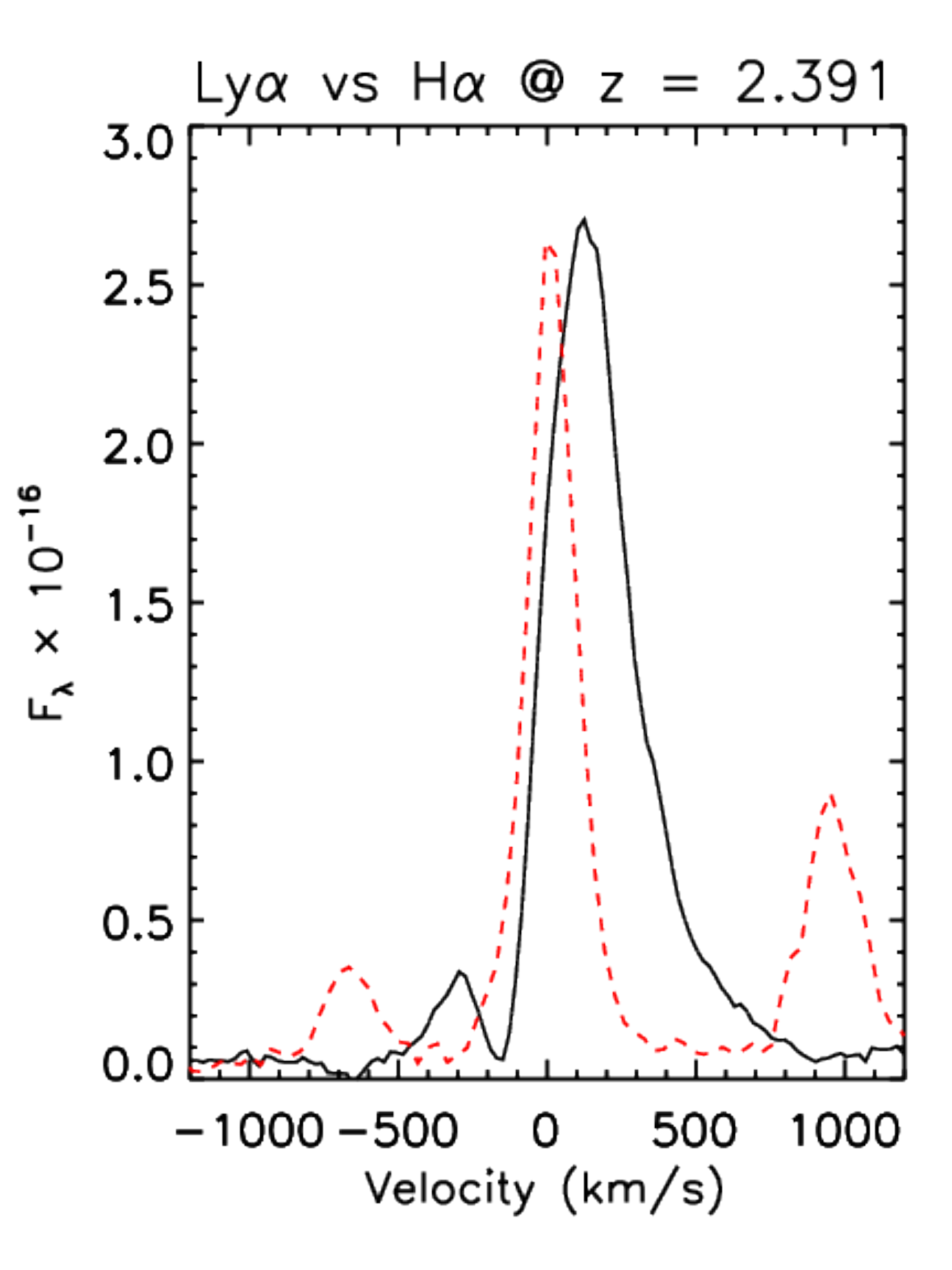}

\caption{Top: Integrated H$\alpha$ flux map from SINFONI spectral fits with continuum-subtracted H$\alpha$ flux contours in black and 
continuum-subtracted Ly$\alpha$ flux contours in blue. Ly$\alpha$ contour levels represent 1$\sigma$, 2$\sigma$, and 3$\sigma$ 
fluxes above the background. White, dashed boxes show regions of SINFONI spaxels that were binned to measure H$\alpha$ flux. Bottom: 
Comparison of H$\alpha$ + [N II]$\lambda\lambda$6548,6584 (red-dashed line) and Ly$\alpha$ (black solid line) emission-line profiles, 
taken from SINFONI and \citet{Rig18} MagE observations, respectively.}
\label{fig:lya_ha}

\end{figure}

We compare the image plane morphology of the H$\alpha$ gas from our SINFONI spectral fits to that of the Ly$\alpha$-emitting gas 
from {\it HST} imaging, as shown in Figure \ref{fig:lya_ha}, to determine if the AGN outflows have some influence on the 
propagation or escape of Ly$\alpha$ photons. We find that the Ly$\alpha$-emitting gas is most prominent between, rather 
than cospatial with, the brightest knots of H$\alpha$ that reside over the AGN and likely star-forming regions. This discrepancy 
between the morphology of Ly$\alpha$ and H$\alpha$ has also been reported in similar studies of local starburst 
galaxies \citep{Ost09,Hay13} and high redshift (z $<$ 2.5) quasar hosts \cite{Bay17}.

We also compare the spectral signatures of Ly$\alpha$ and H$\alpha$ in Figure \ref{fig:lya_ha}, with Ly$\alpha$ emission obtained 
from long-slit MagE observations covering the full spatial extent of the arc as detailed in \citealt{Rig18}. 
Comparable H$\alpha$ emission was obtained by binning SINFONI spaxels that overlap with locations of the strongest Ly$\alpha$ flux 
knots in the {\it HST} imaging (boxes in the flux map of Figure \ref{fig:lya_ha}). The observed velocity structure of Ly$\alpha$ in 
comparison to Balmer emission is typical in studies of green pea galaxies \citep{Yan17,Orl18}. Although the sampled spectra 
are immediately adjacent to the detected AGN outflows, fitting Gaussians to the binned H$\alpha$ spectrum, we measure a FWHM of $\sim$200\,
km s$^{-1}$, which suggests relatively undisturbed kinematics, and do not detect a secondary, outflow component. 
These observations suggest that the AGN outflows in SGAS 0033+02 are anti-correlated with the observed Ly$\alpha$ structure.

To estimate the intrinsic properties of the AGN in the source plane of SGAS 0033+02, we 
must apply a gravitational lens model to our observed image plane data. Details on the methods used 
to convert image plane observations of SGAS 0033+02 into source plane data are detailed in the Appendix. 
From our model, we find that the main arc of SGAS 0033+02 straddles a lensing critical line, which separate 
regions of different image multiplicities. As such, the observed structure in this arc is approximately half 
of the galaxy observed in the counter images. 

\section{Source Plane Analysis}
\label{dis_sec}

\subsection{Extent of AGN Outflows}

\begin{figure}[h]
\centering
\includegraphics[width=0.48\textwidth]{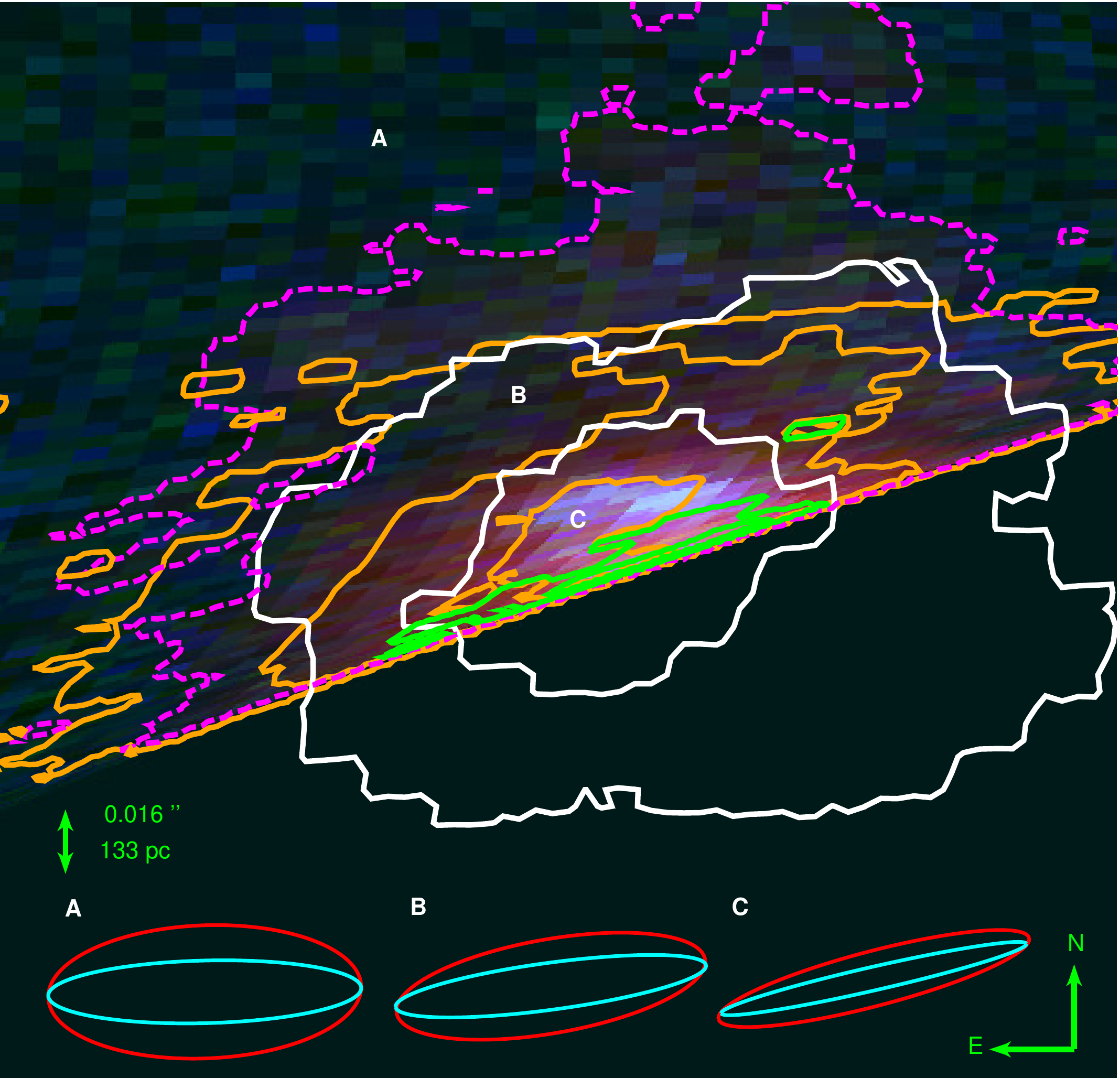}
\caption{Source plane reconstruction of {\it HST} F140W/F814W/F555W WFC3 imaging for the northern half of the main arc. 
Overplotted orange and green contours represent reconstructed narrow- and broad-component H$\alpha$ fluxes from Figure 
\ref{fig:hamaps}, respectively. The magenta and white contours represent rest-frame optical continuum F140W image fluxes 
of the source plane reconstruction for the arc and Counter Image 1, respectively. Source plane images of the arc and 
counter image are aligned manually, as parametric modeling does not match the position in the source plane between images.
Red and cyan ellipses below represent 0.140" and 0.067" circular PSF FWHMs for F140W and F555W filters, respectively, lensed back 
to the source plane. Ellipses are labeled with letters which map the transformation moving toward the caustic line.}
\label{fig:src_rec_src_plan}
\end{figure}

Figure \ref{fig:src_rec_src_plan} shows the source plane reconstruction of the fraction of SGAS 0033+02 observed in the main arc 
as it would have been seen without the presence of the lens. Orange and green contours represent the source 
plane extents of the narrow and broad H$\alpha$ emission-line components from SINFONI observations shown in 
Figure \ref{fig:hamaps}, respectively. Measuring the radial extent of the broad-component (i.e. outflows) in the 
source plane, we report a length of r $\sim$100 pc. This is likely the maximum outflow extent in the observed 
half of the galaxy, as the location of the outflows is adjacent to the rest-frame optical continuum peak of the 
galaxy in Figure \ref{fig:src_rec_src_plan}, which suggests that they reside near the galaxy nucleus and AGN. However, this 
measurement should be treated as a lower limit of the true outflow extent, as we have no kinematic data on the other half of 
the galaxy which is not observed in the arc. We can measure the distance between the furthest knot of emission in the other 
half of the galaxy, as seen in the F555W image of Counter Image 1 which traces the extent of the observed H$\alpha$ emission, 
and its F140W continuum peak to set an upper limit on the maximum possible outflow distance as r $\sim$830 pc.

\subsection{Intrinsic Flux Measurement}

By reconstructing the source plane image of SGAS 0033+02, we can also determine the magnification any given point 
in the image plane. Demagnified fluxes for the AGN-ionized outflows in SGAS 0033+02 were obtained 
by dividing the image plane datacubes by a magnification map at matched pixel scale resolution, as determined 
from our strong lens model. Spectra in the central $0.5" \times 0.5"$ square were again binned and fit to measure 
the total demagnified flux. In this second iteration of fitting, line dispersons and centroids were fixed to the 
values obtained from the fit to the image-plane spectrum, with only the total flux (i.e. the Gaussian amplitude) 
allowed to vary. Source plane fluxes are provided in Table \ref{tab:binflux}.

Before analyzing our measured fluxes, we applied a reddening correction using a standard Galactic reddening curve 
\citep{Sav79} and color excesses calculated from the observed, source-plane H$\alpha$/H$\beta$ ratio \citep{Ost06}, 
assuming an intrinsic recombination value of 2.85. The extinction was calculated using

$$ E(B-V) = -\frac{2.5log(\frac{F_o}{F_i})}{R_{\lambda}} = \frac{2.5log(\frac{(H\alpha/H\beta)_i}{(H\alpha/H\beta)_o})}{R_{H\alpha} - R_{H\beta}} $$

\noindent where E(B-V) is the color excess, R$_{\lambda}$ is the reddening value at a particular wavelength, and 
F$_o$ and F$_i$ are the observed and intrinsic fluxes, respectively. Galactic reddening values are R$_{H\alpha} \approx$ 2.5 
and R$_{H\beta} \approx$ 3.7. Corrected line fluxes are then given by 

$$ F_i = \frac{10^{-0.4~R_{\lambda}~E(B-V)}}{F_o} $$

\noindent with dereddened source-plane fluxes listed in Table \ref{tab:binflux}.

\subsection{AGN Mass Outflow Rate}

We use the dereddened source-plane flux of the broad H$\alpha$ component to 
estimate the total, observed ionized gas mass in the NLR outflows, for case B recombination \citep{Pet97,Ost06}. 
The total luminosity of H$\beta$, originating from clouds within a total volume V$_c$, is $L(H\beta) =
n_{e}n_{p}a^{eff}_{H\beta}h\nu_{H\beta}V_{c}$, with $a^{eff}_{H\beta}$ and $\nu_{H\beta}$ 
being the effective recombination coefficient and rest frequency of H$\beta$, and $n_{e}$ 
and $n_{p}$ being the number densities of electrons and protons. We assume completely 
ionized hydrogen clouds, therefore $n_{e} \sim n_{p}$. H$\alpha$ and H$\beta$ luminosities 
are related such that $L(H\alpha) = (j_{H\alpha}/j_{H\beta})L(H\beta)$, where 
$j_{H\alpha}/j_{H\beta}$ is the intensity of H$\alpha$ relative to H$\beta$. 
Assuming the same density for all clouds, $n_{p}m_{p}$, with $m_{p}$ being the proton mass, 
the total ionized gas mass is $M_{NLR} = n_{p}m_{p} \times V_c$. From the relations made above:

$$ M_{NLR} = \frac{m_{p}L(H\alpha)}{n_{e}(j_{H\alpha}/j_{H\beta})a^{eff}_{H\beta}h\nu_{H\beta}}$$

$$ M_{NLR} = 2.523\times 10^5 \times L_{42}(H\alpha) M_{\astrosun}$$

\noindent where $L_{42}(H\alpha)$ is in units of 10$^{42}$\,erg s$^{-1}$. 
Intrinsic $a^{eff}_{H\beta}$ and $j_{H\alpha}/j_{H\beta}$ were taken from \citet{Ost06}, assuming a 
temperture of $T = 10^4$\,K. We derive an electron density (n$_e$ cm$^{-3}$) using an estimated [S II] 
$\lambda\lambda$6716/6731 line ratio for the AGN ionized broad-emission-line component \citep{All79,Ost06}. 
We measure a dereddened [S II] $\lambda$6731 broad-component flux of 3.13$\times$10$^{-18}$\,erg s$^{-1}$ cm$^{-2}$, 
do not detect a comparable broad [S II] $\lambda$6716 component, and instead use a flux of 1.61$\times$10$^{-18}$\,erg 
s$^{-1}$ cm$^{-2}$ as a flux upper limit as this represents a dereddened 3$\sigma$ flux detection at this 
wavelength, assuming a similar line disperson. These fluxes produce a maximum ratio $\sim$ 0.5, from which we assume 
$n_e \sim 10^4$\,cm$^{-3}$. Using a luminosity distance of $D_L$ = 6.071$\times$10$^{28}$\,cm \citep{Wri06}, 
we measure the dereddened L(H$\alpha$) of the outflowing wind to be 2.92$\times 10^{41}$\,erg s$^{-1}$ in the 
source plane, and calculate a gas mass of $M_{NLR} = 7.37 \times 10^4 M_{\astrosun}$. As this measurement is 
derived from a 0.5$" \times$0.5$"$ bin containing spectra from both sides of the critical line, 
the reported value assumes similar fluxes on the side of the lensed system hidden by the lensing critical line.
Using this gas mass, we then calculate the mass outflow rate in this region by dividing the total 
mass $M$ by the time $t$ it takes to travel across the extent which we observe the outflows, where $t = R/v$. 
We assume a maximum outflow extent of $\sim$ 100 pc, as derived from the strong lens model. Observed 
radial velocities of the outflows are on the scales of tens of km s$^{-1}$, however, these are likely compromised 
by projection effects. We instead use the maximum blueshifted velocity defined as 1/2 $\times$ the full width 
at zero maximum (FWZM), approximately the 3$\sigma$ velocity offset from the centroid of the broad H$\alpha$ 
component measured in our binned spectrum as our deprojected velocity, $v = 924$\,km s$^{-1}$. Using these 
parameters, we measure a mass outflow rate of $\dot{M} =$ 0.67\,M$_{\astrosun}$\,yr$^{-1}$. The power of 
the outflow $dE/dt$ is then calculated as:

$$\frac{dE}{dt} = 0.5\frac{Mv^2}{t} $$

\noindent for a log($\dot{E}$) = 41.33\,erg s$^{-1}$. We use the dereddened source-plane flux of the [O~III] 
$\lambda$5007 and [O I] $\lambda$6300 broad components to measure the bolometric luminosity of the AGN, using the 
method from \citealt{Net09},
log (L$_{bol}$) =  3.8 + 0.25 logL([O~III]$\lambda$5007) + 0.75 logL([O I]$\lambda$6300), for a log(L$_{bol}$) = 45.02\,erg s$^{-1}$. 
The resulting ratio of outflow power to bolometric luminosity is log($\dot{E}/L_{bol}$) = -3.76, 
less than the 0.5\% threshold typically required to provide a significant impact on the host galaxy \citep{Hop10}.

\subsection{Star Formation Rate}

We convert the narrow-line, non-AGN ionized H$\alpha$ luminosity not attributed to AGN ionization (i.e. the H$\alpha$ flux 
measured in the left column of Figure \ref{fig:hamaps}) to a star formation rate (SFR), by using the 
relation from \citet{Ken98}, where SFR $(M_{\astrosun}~yr^{-1}) = 7.9\times10^{-42}L(H\alpha)$, and adjusting to the 
initial mass function (IMF) from \citet{Cha03}, which reduces the SFR by a factor of 1.7. We measure source-plane H$\alpha$ 
luminosities north and south of the lensing caustic to be L(H$\alpha$)$_{north}$ = 4.9$\times$10$^{42}$\,erg s$^{-1}$ and 
L(H$\alpha$)$_{south}$ = 1.5$\times$10$^{43}$ erg s$^{-1}$, which convert to SFRs of 22.8\,M$_{\astrosun}$ yr$^{-1}$ and 
70.7\,M$_{\astrosun}$ yr$^{-1}$, respectively. We note that these rates may be upper limits, as there may be 
contributions to the HII regions by AGN ionization. We can compare our SFR measurements to those in \citet{Liv15}, which show a 
correlation between SFR in star-forming clumps and their sizes, by isolating a lower-limit SFR in the discrete, fully 
imaged H$\alpha$ knot north of the continuum peak, as shown in Figure \ref{fig:hamaps}
and \ref{fig:lya_ha}. Here, we measure a demagnified F(H$\alpha$) = 1.132$\times$10$^{-17}$\,erg s$^{-1}$ cm$^{-2}$, which 
converts to a SFR of 2.43\,M$_{\astrosun}$ yr$^{-1}$, over an area of 0.7\,milliarcsecond$^2$ in the source plane 
(Figure \ref{fig:src_rec_src_plan}) for an approximate radius of 15\,milliarcseconds, or 125\,pc. Measurements for the global 
star formation rate and the clump star formation rate both exceed the mass outflow rate of the AGN. Therefore, the central 
AGN, in its current state, is incapable of displacing enough material to quench star formation in this galaxy.

\section{Discussion}

Producing this spatially resolved analysis of AGN outflows in a "normal" star-forming galaxy at z $\sim$ 2, we find it to 
be similar to weak AGN with strong star formation in the nearby universe. The measured source-plane bolometric luminosity of this 
object suggests that we are observing a Seyfert-like AGN in SGAS 0033+02. In addition, the observed recombination emission-line dispersons 
indicate that SGAS 0033+02 is a Type 2 AGN, where the central engine is obscured along our line of sight. With the observed morphology 
of the bright, outflowing NLR being relatively compact, we find this target to be analogous to the nearby Seyfert 2 NGC 1068 
\citep{Cre00c,Das06}. From {\it HST} WFPC2 [O~III] imaging \citep{Sch03}, the enclosed [O~III] flux within a 100 pc radius of the 
nucleus for this nearby AGN is 9.11 $\times 10^{-12}$\,erg s$^{-1}$ cm$^{-2}$, which converts to a luminosity of L([O~III]) = 
1.7$\times$10$^{41}$\,erg s$^{-1}$. The measured flux in this system originates from one half of the NLR, with the other half extinguished 
below the plane of its host disk. For comparison, we can divide the measured L([O~III]) of SGAS 0033+02 of 4.32$\times$10$^{41}$\,erg s$^{-1}$ in 
half and find its luminosity to be on par with NGC 1068 at L([O~III]) = 2.16$\times$10$^{41}$\,erg s$^{-1}$. Notably, as a Seyfert AGN 
with an intrinsic observed F([O~III]) = 3.87$\times$10$^{-18}$\,erg s$^{-1}$ cm$^{-2}$, it is unlikely that the broad 
emission-line component attributed to AGN ionization in SGAS 0033+02 would be detected in a typical field galaxy at z = 2.391. 
Combined with the effects of star-formation dilution hiding narrow AGN NLR signatures near systemic velocity, these findings suggest 
many AGN may go undetected in surveys of galaxies residing near Cosmic Noon \citep{Tru15}. 


There is no evidence in our current observations that we are missing broad outflowing emission-line components at greater radii 
due to lesser amounts of magnification. As shown in Figure \ref{fig:lya_ha}, binning H$\alpha$ lines exterior to where we detect 
AGN outflows result in single Gaussian fit without the presence of a second, broad component. However, it remains unclear if we are 
observing the true extent of the AGN outflows because, as described above, the main arc of SGAS 0033+02 is a partial image that contains 
roughly half of the galaxy seen in the counter images. Assuming that the outflows originate from the optical continuum peak, we cannot 
know the extent of the winds in the other half of the system without kinematics measurements for one of the counter images. 

Comparing the extent of the AGN-ionized region to the AGN [O~III] luminosity of SGAS 0033+02, we find that it has a relatively 
small extent for its luminosity when compared to the radius vs luminosity correlation for NLRs found in previous studies 
\citep{Sch03,Liu10,Fis18,Dem18}. Although we are likely observing the maximum extent of the AGN outflows in our observations, 
a narrow AGN-ionized emission-line component displaying rotation kinematics that extends to larger distances would be undetected 
due to dilution by the larger flux contribution of the HII star forming region. Measuring the source plane radial 
extent of the [O~III] emission shown in Figure \ref{fig:hamaps}, we find a maximum R$_{[O~III]}$ $\sim$ 800 pc. Assuming an 
AGN-ionized component exists throughout, a radial extent of 800 pc paired with a log(L[O~III]) = 41.6\,erg s$^{-1}$ 
places SGAS 0033+02 in line with previous findings from Seyferts and QSOs in the nearby universe.


\section{Conclusions}
\label{conc_sec}

We have analyzed spatially-resolved, rest-frame UV / optical imaging and spectroscopy of a Seyfert AGN at z $\sim$ 2 for the first time. 
Our major findings are:

1) AGN-ionized outflows extend to a radius of r\,$\sim$\,100 pc. We calculate a mass outflow rate over this distance of 
$\dot{M} = 0.55\,M_{\astrosun}~yr^{-1}$. The corresponding ratio of outflow power to bolometric luminosity is exceedingly low, 
log($\dot{E}/L_{bol}$) = -3.76, suggesting the AGN does not significantly impact the host galaxy.

2) SGAS 0033+02 also exhibits a star formation rate on the order of tens of solar masses per year, which greatly 
exceeds the AGN mass outflow rate. As such, the current state of the AGN in SGAS 0033+02 would be unlikely to quench 
star-formation within the galaxy.

3) The positions of outflowing winds and Ly$\alpha$ emission are anti-correlated. Ly$\alpha$ exists where the outflow is not,
therefore the outflow has not destroyed Ly$\alpha$ over the whole arc. Ly$\alpha$ structure in this galaxy is also similar 
to those in galaxies not hosting AGN.

4) SGAS 0033+02 resembles weak AGN with strong star formation observed in the local universe. Faint emission-line 
signatures of these low-luminosity AGN make their detection at z$\sim$2 extremely difficult without gravitational lensing. 
Combining faint AGN emission with line-dilution from strong star formation, it is possible that many AGN are 
missed in survey work at this redshift.

\acknowledgments

Based on observations collected at the European Organisation for Astronomical Research in the Southern Hemisphere 
under ESO programmes 094.A-0746(A) and 098.A-0459(A).

Based on observations made with the NASA/ESA Hubble Space Telescope, obtained from the data archive at the Space 
Telescope Science Institute. STScI is operated by the Association of Universities for Research in Astronomy, Inc. 
under NASA contract NAS 5-26555.

This paper includes data gathered with the 6.5 meter Magellan Telescopes located at Las Campanas Observatory, Chile.

The authors would like to thank the anonymous referee for their helpful comments. This paper benefited from error analyses 
by M. Revalski and discussions with D. M. Crenshaw and C. L. Gnilka. TCF was supported by an appointment to the NASA 
Postdoctoral Program at the NASA Goddard Space Flight Center, administered by Universities Space Research Association 
under contract with NASA and by NASA through grant number HST-AR-15019.006-A from the Space Telescope Science Institute, 
which is operated by AURA, Inc., under NASA contract NAS 5-26555. SL was partially funded by UCh/VID project ENL18/18. 
LFB was supported by Anillo ACT-1417.


\bibliographystyle{apj}             
\bibliography{apj-jour,sgas0033}       

\appendix
\section{Gravitational Lens Modeling}
\label{sec:modeling}

\subsection{Lensing Mass Models Methodology}
\label{sec:Methodology}

Here, we provide a brief summary of the gravitational lensing analysis used in this work and 
we refer the reader to \cite{Kneib1996}, \cite{Smith2005}, \cite{Verdugo2011} and \cite{Richard2011} for a more in depth 
description. We take a parametric approach, using {\tt{Lenstool}} \citep{Jullo2007} to model the cluster 
mass distribution surrounding our target as a series of dual pseudo-isothermal ellipsoids (dPIEs, \citealt{Eliasdottir2007}), 
which are optimized through a Monte Carlo Markov Chain minimization. 

To model the cluster mass distribution, Dark Matter (hereafter DM) dPIE clumps are combined to map the DM at the cluster scale. 
Galaxy scale DM potentials are used to describe galaxy scale substructure. Considering the large number of galaxies in the cluster, 
it is not feasible to optimize the parameters of every potential, as the large parameter space will lead to an unconstrained 
minimization. Moreover, individual galaxies contribute only a small fraction to the total mass budget of the cluster, so their 
effects on lensing are minimal unless they are in close proximity, in projection, to the lensed galaxies. To reduce the 
overall parameter space we scale the parameters of each galaxy to a reference value, using a constant mass-luminosity scaling 
relation (see \citealt{Limousin2007}).

\subsection{Selection of Cluster Members}
\label{sec:cm}

We used Sextractor in the "white" image of the MUSE data to detect all the sources and define apertures for 
PyMuse (https://pypi.org/project/PyMUSE/) to integrate the spaxels and thus to obtain the spectra for each of the galaxies. 
PyMuse can also run Redmonster (Hutchinson et al. 2016) to determine individual redshifts. All the spectra, and Redmonster 
best candidates, were visually inspected to assign the redshift for each galaxy.

We then constructed a galaxy cluster catalog using the red sequence technique \citep{Gladders2000}, by selecting in a color-magnitude 
diagram galaxies that show a similar color. Our final catalog contains 80 cluster members.

As the brightest galaxies, or bright cluster galaxies (BCGs), of galaxy clusters tend to not follow the cluster red sequence, 
we remove the BCG of the South-East sub cluster \citep{Newman2,Newman1}. We keep the other BCG 
in the scaling relation due to the lack of constraints to properly model the lensing potential shape on that side.
In addition, we detected several spirals galaxies in the MUSE data cube at z$\sim1.03$ (Barrientos et al. in prep.) 
that may influence the lensing configuration of the bright arc of SGAS 0033+02. We include the two closest ones 
($\alpha=00^{\rm h} 33^{\rm m} 41.6576^{\rm s}$, $\delta=+02^{\rm o} 42\arcmin 13.7186\arcsec $ and 
$\alpha=00^{\rm h} 33^{\rm m} 41.0841^{\rm s}$, $\delta=+02^{\rm o} 42\arcmin 05.5126\arcsec $) in our lensing 
potentials, but model them separately as individual potentials at the cluster redshift.

\subsection{Lensing Constraints}
\label{sec:cstr}

\begin{figure*}[h]
\centering
\includegraphics[scale=0.6]{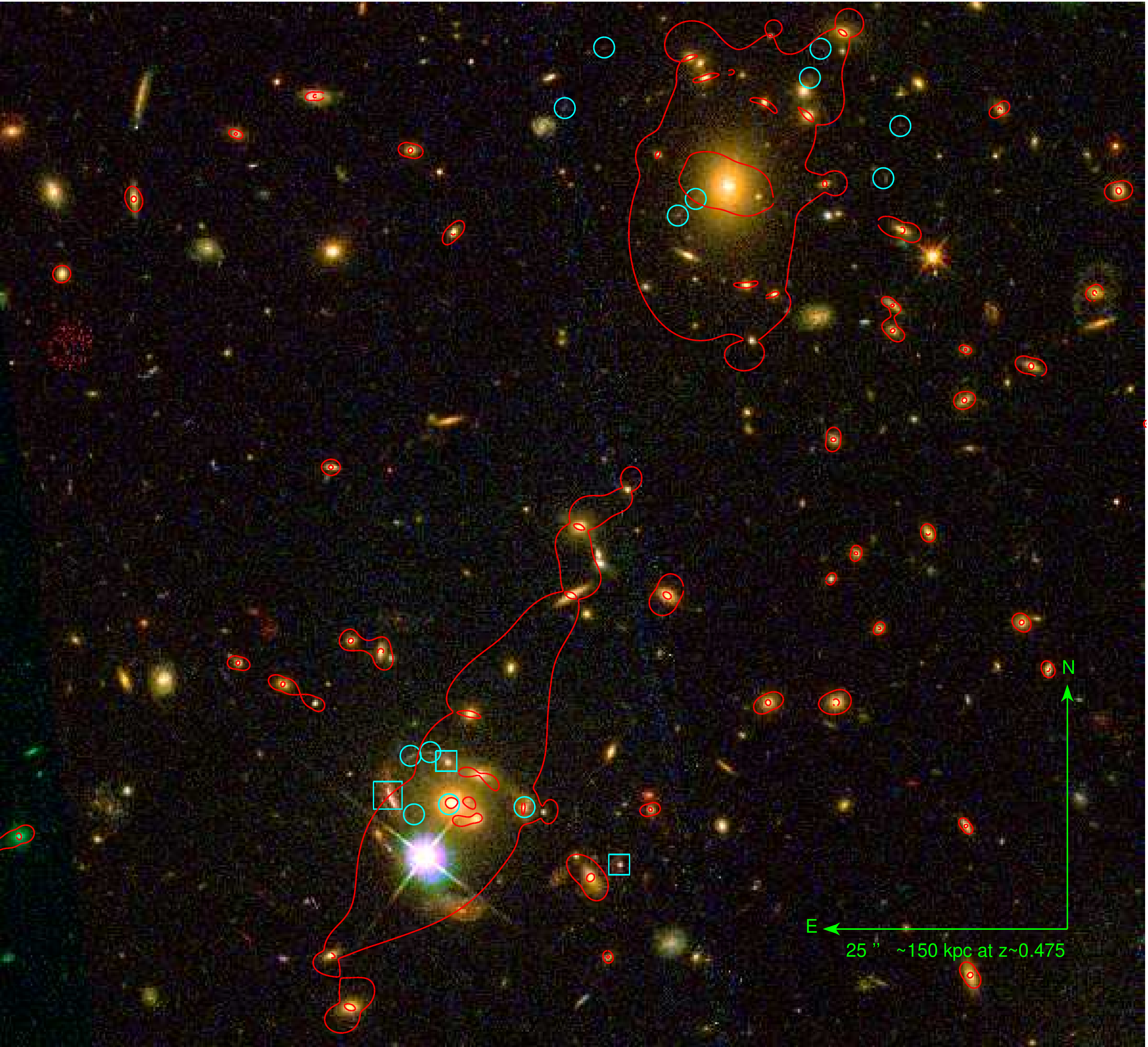}

\caption{{\it HST} F555W/F814W/F140W image detailing the locations of all the strong lensing observed in the field. 
The red curve is the critical line at z$=$2.39, the redshift of SGAS 0033+02. Cyan circles are constraints use in the model. Cyan squares 
are the region where the SGAS 0033+02 constraints are positioned.}
\label{fig:cluster}

\end{figure*}

\begin{figure*}[h]
\centering
\includegraphics[scale=0.6]{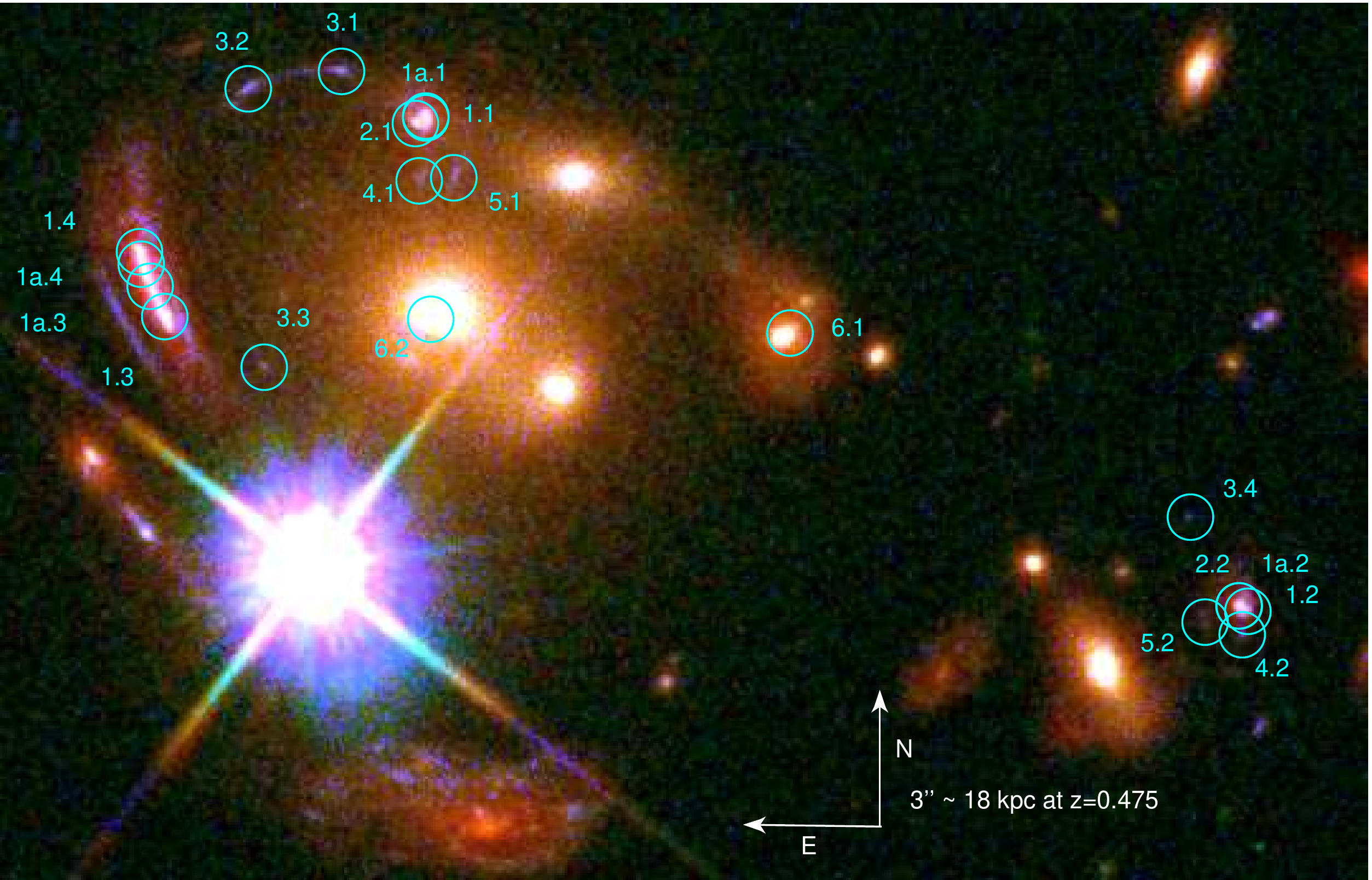}
\caption{Zoom in on the southern cluster in Figure \ref{fig:cluster} identifying individual contraints in the lensing model as cyan circles. 
System 1 and 1a are almost overlapping, see Table \ref{tab:mul} for more information.}
\label{fig:subcluster}  
\end{figure*}

We consider a large number of constraints for the bright arc in order to obtain the most accurate 
source reconstructions. Figure \ref{fig:cluster} exhibits an {\it HST} F555W/F814W/F140W image 
marking the positions of all constraints used in our model and the resultant critical line. We also provide 
an enlarged, labeled image of the region near SGAS 0033+02 in 
Figure \ref{fig:subcluster}, with the positions and redshifts of these systems listed in Table \ref{tab:mul}. 
From our model, we find the lensing critical line at z = 2.39 lies directly over the center of the arc of SGAS 0033+02, 
such that the north and south ends of the arc are reflections of one another. This is supported by the symmetries 
on each side of the arc observed both in imaging and kinematics.

\begin{figure}[h]
\centering
\includegraphics[scale=0.45]{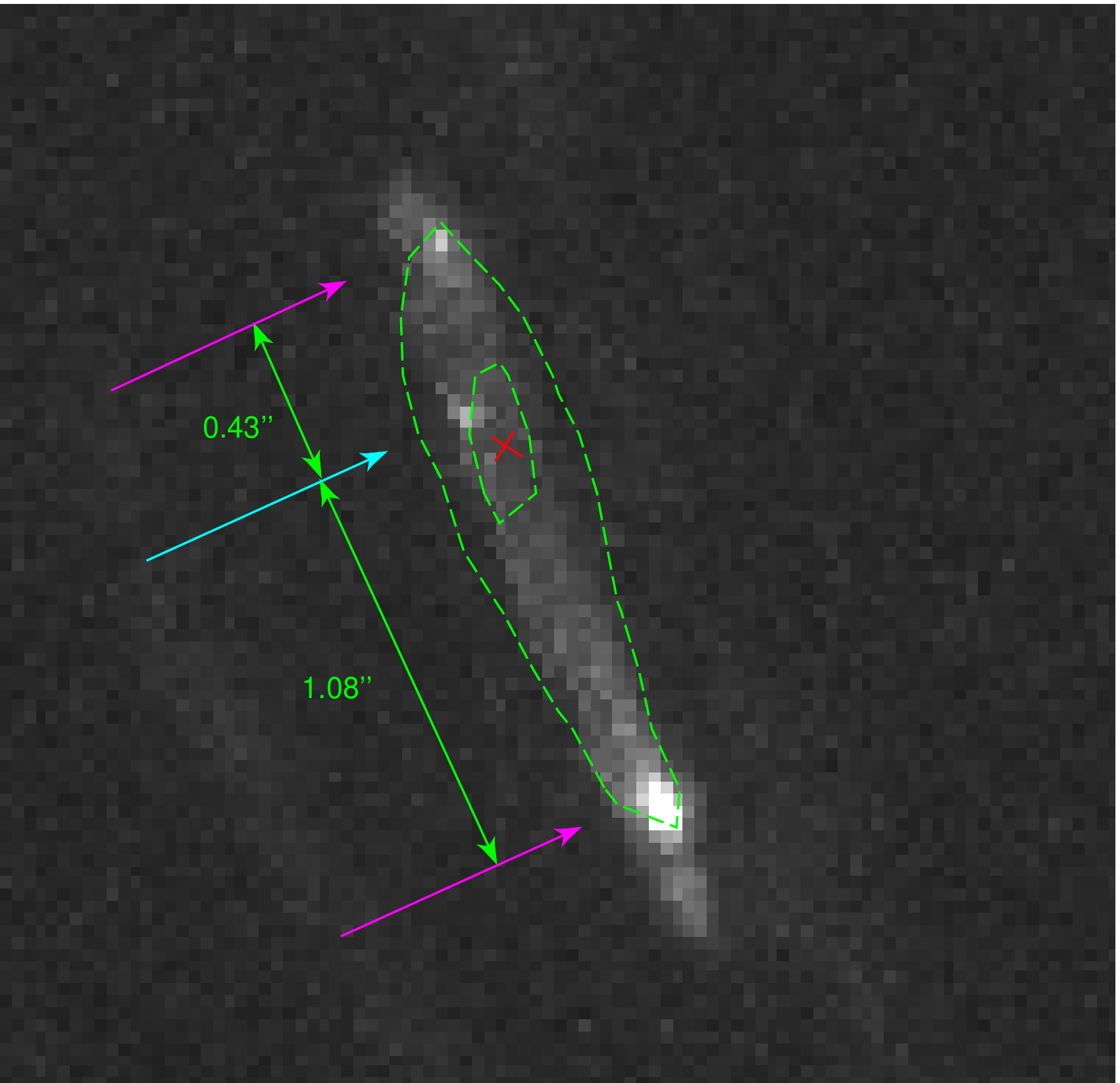}

\caption{{\it HST} F555W image showing the asymmetry is the SGAS 0033+02 arc morphology. The green dashed contour shows 
the luminosity contour of the F140W HST band. The red cross shows the rest-frame optical continuum peak and expected 
critical curve crossing. The two magenta arrows show the two bright knots that are identified as being the same emission 
knot on either side of the critical line. The cyan arrow marks the bright spot that does not have a symmetrical counterpart.}
\label{fig:asym}

\end{figure}

We find that the arc contains an unusual asymmetry that cannot be accounted for by the lensing model, 
observed in the rest-frame UV continuum F555W image, as shown in Figure \ref{fig:asym}. 
As the critical line from the strong lensing model crosses at the flux peak in the F140W image, we observe 
that the small and faint emission knot just north to the critical curve does not show a symmetric counterpart 
on the other side of the arc. As such, this emission could be due to a transient in the arc and we do not include 
this feature in our constraints. In addition, the southern emission knot in the F555W image is significantly 
brighter than the corresponding knot in the top arc. This knot coincides with the H-alpha knot visible in SINFONI 
data (see Figure \ref{fig:hamaps}), which also show this flux asymmetry. A possible explanation to this discrepancy 
is that the observed flux of this feature is time variable, however additional observations are required to test 
such a scenario.  

\begin{table}[h]
\centering
\caption{lensing constraints}
\label{tab:mul}
\begin{tabular}{llllcc}
\hline 
ID & $\Delta\alpha^{\rm ~a}$ & $\Delta\delta^{\rm ~a}$ &  z$_{spec}^{\rm b}$ & z$_{model}^{\rm c}$ & rms$^{\rm d}$\\ 
 & h:m:s & d:m:s  &  & (\arcsec)\\ 
\hline 
1.1 & 0:33:41.167 & +2:42:21.200 &  2.39 &  --& 0.15\\
1.2 & 00:33:39.977 & +02:42:10.4602 & 2.39 &  -- &  0.08\\ 
1.3 & 00:33:41.549 & +02:42:16.802 &  2.39 &  --&  0.20\\ 
1.4 & 00:33:41.586 & +02:42:18.223 &  2.39 &  --&  0.13\\ 
1a.1 & 00:33:41.167 & +02:42:21.1589 &  2.39 &  --& 0.14\\ 
1a.2 & 00:33:39.9765 & +02:42:10.4568 &  2.39 &  -- & 0.08\\ 
1a.3 & 00:33:41.5704 & +02:42:17.4662 &  2.39 &  --& 0.09\\ 
1a.4 & 00:33:41.5835 & +02:42:17.9503 &  2.39 &  --& 0.09\\ 
2.1 & 00:33:41.1822 & +02:42:21.0352 &  2.39 &  --& 0.09\\ 
2.2 & 00:33:39.964 & +02:42:10.3392 &  2.39 &  --&  0.15\\ 
3.1 & 00:33:41.2903 & +02:42:22.1782 &  2.096 &  --& 0.35\\ 
3.2 & 00:33:41.4269 & +02:42:21.7955 &  2.096 &  --& 0.05\\ 
3.3 & 00:33:41.4035 & +02:42:15.6917 &  2.096 &  --& 0.46\\ 
3.4 & 00:33:40.048 & +02:42:12.4045 &  2.096 &  --&  0.05\\ 
4.1 & 00:33:41.1767 & +02:42:19.7782 &  2.39 &  --&  0.05\\ 
4.2 & 00:33:39.9723 & +02:42:09.8256 &  2.39 &  --&  0.09\\ 
5.1 & 00:33:41.126 & +02:42:19.845 & 2.39 &  --& 0.11\\ 
5.2 & 00:33:40.0265 & +02:42:10.1024 &  2.39 &  --& 0.30\\ 
6.1 & 00:33:40.634 & +02:42:16.441 &  0.969 &  --& 0.04\\ 
6.2 & 00:33:41.1598 & +02:42:16.7427 &  0.969 &  --& 0.10\\ 
7.1 & 00:33:38.019 & +02:43:27.487 &  -- &  $5.70^{+0.69}_{-0.54}$ & 0.12\\ 
7.2 & 00:33:38.574 & +02:43:35.552 &  -- &  $5.70^{+0.69}_{-0.54}$ & 0.36\\ 
7.3 & 00:33:40.08 & +02:43:35.6793 &  -- &  $5.70^{+0.69}_{-0.54}$ & 0.21\\ 
8.1 & 00:33:39.4425 & +02:43:19.8843 &  -- &  $5.72^{+2.03}_{-0.09}$ & 0.06\\ 
8.2 & 00:33:39.5671 & +02:43:18.1456 &  -- &  $5.72^{+2.03}_{-0.09}$ & 0.05\\ 
9.1 & 00:33:38.137 & +02:43:22.043 &  -- &  $3.92^{+0.32}_{-0.34}$ &  0.09\\ 
9.2 & 00:33:38.648 & +02:43:32.52 &  -- &  $3.92^{+0.32}_{-0.34}$ &  0.11\\ 
9.3 & 00:33:40.3541 & +02:43:29.3617 &  --& $3.92^{+0.32}_{-0.34}$  &   0.06\\ 
\hline 
\end{tabular}
\medskip\\
$^{\rm a}$ $\Delta\alpha$ and $\Delta\delta$ are the position of the arc.~~~~~~~~~~~~~~~~~~~~~~~~~~~~~~~\\[1pt]
$^{\rm b}$ z$_{spec}$ refer to the spectroscopic redshift set to the arc, when no value are presents we let the redshift free to vary during the minimization. ~~~~~~~~~~~~~~~~~~~~~~~~~~~~~~~\\[1pt]
$^{\rm c}$ z$_{model}$ refer to the best redshift found by our modelisation. ~~~~~~~~~~~~~~~~~~~~~~~~~~~~~~~\\[1pt]
$^{\rm d}$ The rms refers to the square root of the mean square of the predicted image position and represents the goodness of our fiducial model regarding the position of each constraint. ~~~~~~~~~~~~~~~~~~~~~~~~~~~~~~~\\[1pt]
\end{table}

\begin{figure*}[h]
\centering
\includegraphics[scale=0.7]{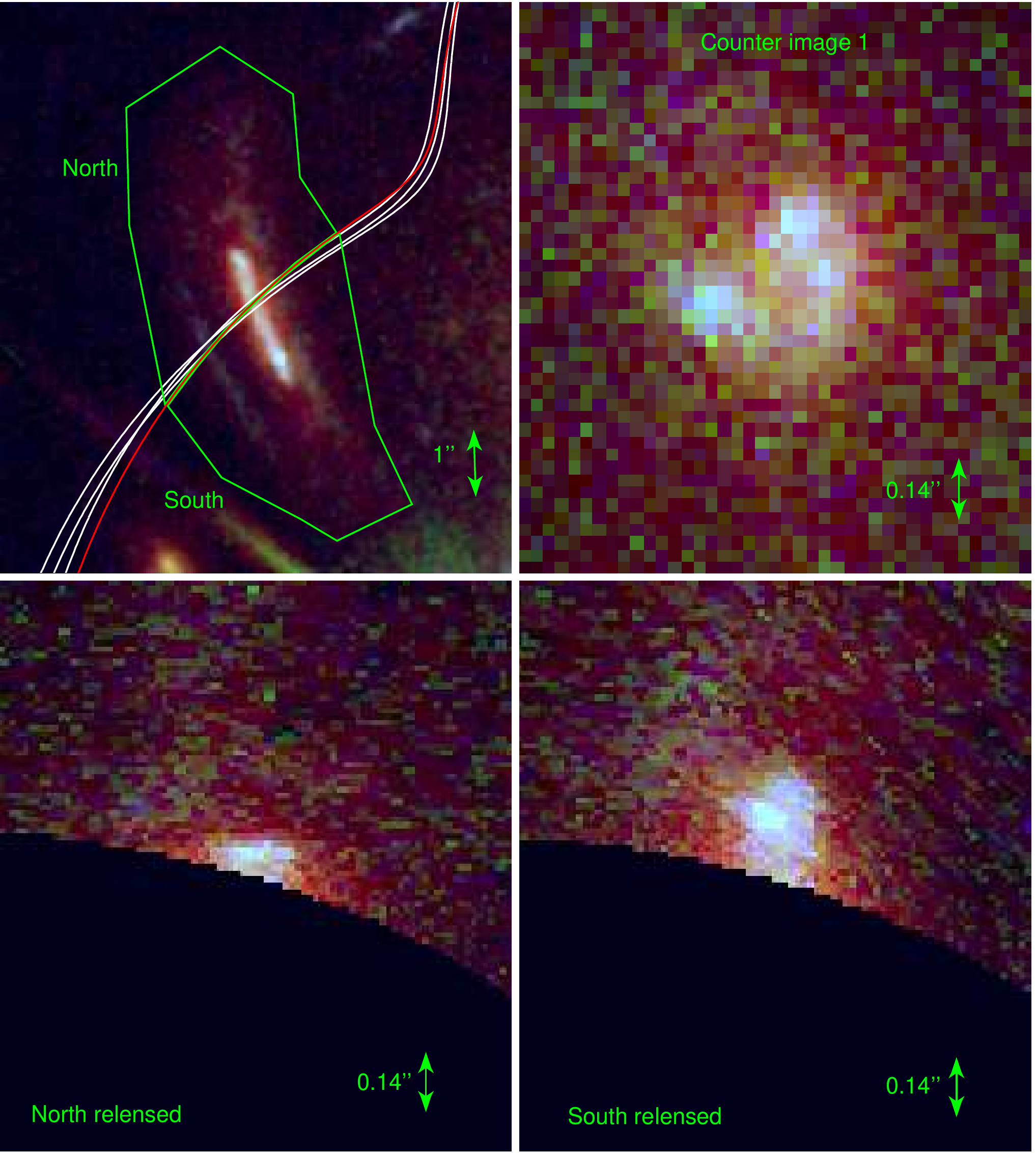}

\caption{Top: Zoom in of Figure \ref{fig:cluster} on the main arc (left) and Counter Image 1 (right) of SGAS 0033+02. 
The main arc image of SGAS 0033+02 shows the location of the critical line of our lensing model and green polygon regions north 
and south of the critical line which are used to produce the reconstructed source plane images. 
Bottom left : The relensed image of the North bin from the top left image matched to the position of 
Counter image 1. Bottom right : The relensed image of the South bin from the top left 
image matched to the position of Counter image 1.}
\label{fig:src_rec_img_plan}

\end{figure*}

\end{document}